\documentclass[12pt]{iopart}

\expandafter\let\csname equation*\endcsname\relax
\expandafter\let\csname endequation*\endcsname\relax

\usepackage{diagbox}
\usepackage[table]{xcolor}
\usepackage{iopams}       
\usepackage{amsmath}      
\usepackage{amsfonts}     
\usepackage{amssymb}      
\usepackage{mathrsfs}     
\usepackage{mathtools}    
\usepackage{multirow}     
\usepackage{hyperref}     
\usepackage{tikz}         
\usepackage{graphicx}     
\usepackage{subcaption}   

\begin{document}

\title[]{Emergence and tunability of Fermi-pocket and electronic instabilities in layered Nickelates}

\author{Alpesh SHETH$^1$,Claudine LACROIX$^2$,Sébastien BURDIN$^1$}

\address{$^1$ Université de Bordeaux, CNRS, LOMA, UMR 5798, 33400 Talence, France.\\
$^2$ Institut Néel, CNRS and Université Grenoble-Alpes, Boite Postale 166, 38042 Grenoble Cedex 09, France.}
\ead{alpeshsheth.phy@gmail.com}
\vspace{10pt}
\begin{indented}
\item[]
\end{indented}

\begin{abstract}

Layered Nickelates have gained intensive attention as potential high-temperature superconductors, showing similarities and subtle differences to well-known Cuprates. This study introduces a modelling framework to analyze the tunability of electronic structures by focusing on effective orbitals and additional Fermi pockets, mimicking doping or external pressure qualitatively. It investigates the role of the $3d_{z^2}$ orbital in interlayer hybridization, which leads to the formation of a second pocket in the Fermi surface. The resulting effective model also predicts specific charge and spin susceptibility in the form of Lindhard susceptibility at wave vector $\mathbf{q_{0}} = (\pi, \pi)$, which can be tuned by doping or pressure. These results provide valuable insights into tunable orbital contributions and their influence on potential ordering and electronic instabilities in Layered Nickelates.

\end{abstract}

\section{Introduction}
The discovery of superconductivity in hole-doped layered Nickelates such as $RNiO_2$ (R = La, Nd, Pr) \cite{Zeng2020Oct, Zeng2022Feb, Osada2020Dec, Osada2021Nov, Li2019Aug, Li2020Jul} has led to extensive studies into their electronic and magnetic properties and potential pairing mechanisms. These studies often compare layered Nickelates with relatively well-studied Cuprates. Despite 0similar structure and isoelectronic configurations of $Ni^{1+}:3d^{9}$, layered Nickelates have subtle distinctions compared to Cuprates, such as dependence on the synthesis method, the role of 3-dimensionality \cite{LaBollita2022Sep}, debatable charge order \cite{Tam2022Oct,Parzyck2024Apr}, questionable magnetic behavior \cite{Zeng2020Oct, Li2019Aug,Hu2019Dec, Fowlie2022Sep, Shi2024Feb,Rossi2021Dec, Jiang2020May,Hepting2020Apr}, and the possible presence and origin of a second pocket in Fermi surface \cite{Ding2024Mar, Sun2024Mar,Gu2020Nov, Sakakibara2020Aug, Jiang2019Nov, PhysRevX.11.011050, Botana2022Feb, Lechermann2020Oct, Lechermann2022Apr}.

Typically, studies of these distinctions focus on the superconductivity of layered Nickelates or emphasize the differences between layered Nickelates and Cuprates. Rarely do they explore how specific synthesis steps, such as removing apical Oxygen, layer formation, doping, or applying external pressure, affect the electronic structure and instabilities. One alternative approach to understand these distinctions is to examine the changes in electronic structure and instabilities during the stoichiometric reduction from its parent compound, the perovskite Nickelates $(RNiO_{3})$. In a broader context, the parent perovskite Nickelates themselves are known for their rich electronic and magnetic properties, including metal-insulator transitions (MIT), charge ordering, novel magnetic phases, and bond disproportionation  \cite{Medarde1997, Piamonteze2002Jul, ALONSO199718, Garcia-Munoz1994Jul, Bisogni2016Oct, Varignon2017Apr, Mizokawa1991Sep}. However, the specific mechanisms driving those aforementioned phenomena in perovskite Nickelates are still under active investigation \cite{Bisogni2016Oct,Varignon2017Apr}. Furthermore, perovskite \cite{Yamamoto2002May,Lu2017May,Kim2020Mar} and layered Nickelates \cite{Krieger2022Jul,Ren2023Mar,Zhang2024Aug,Lane2023May,Zhang2023Oct} exhibit competing electronic instabilities confirmed by various experimental and theoretical studies, but there are no reports of superconductivity in perovskite Nickelates.

Scrutinizing the orbital contribution in perovskite and layered Nickelates, Angle-Resolved Photoemission Spectroscopy (ARPES) studies on $RNiO_{3}\,\,(R=La, Nd)$ reveal the $3d_{x^2-y^2}$ and $3d_{z^2}$ orbitals to be the primary dominant contributors to the Fermi surface in perovskite Nickelates \cite{Dhaka2015Jul,King2014Jun}. Meanwhile, ARPES measurements \cite{Ding2024Mar, Sun2024Mar, DiCataldo2024May, Si2024Aug} of layered Nickelates show contribution only from the $3d_{x^2-y^2}$ orbital. On the other hand, several DFT-based tight-binding models on layered Nickelates have captured two pockets in the Fermi surface  \cite{Lechermann2020Oct, Gu2020May, Chen2023Sep, PhysRevX.10.011024, Lee2004Oct, Plienbumrung2022Oct, Klett2022Feb, Zhang2020Feb, Kitatani2020Aug, Hepting2020Apr, Xie2022Jul, Jiang2019Nov}. The orbital origin of one of these two pockets is consistently recognized to be the $3d_{x^2-y^2}$ orbital, while the presence and origin of the second pocket is a topic of non-consensus. Such non-consensus is also present in the potential role of order (magnetic or charge) \cite{Tam2022Oct,Parzyck2024Apr,Krieger2022Jul} in layered Nickelates. On the theoretical side, there are attempts to include interaction effects in layered Nickelates that range from Mott, Hubbard, and Hund physics \cite{Botana2022Feb, Gu2022Jan, Nomura2022Mar, Wang2024Mar, Wang2020Oct,Zhang2020Jan, Wu2020Feb}. These approaches typically use methods beyond Density Functional Theory (DFT) \cite{PhysRevB.101.020503,Hepting2020Apr, PhysRevX.10.011024, Nomura2019Nov}, such as Dynamical Mean Field Theory (DMFT) \cite{Kitatani2020Aug,Lechermann2020Oct} and Dynamical Vertex Approximation \cite{Kitatani2020Aug,Held2022Jan}. We recognize the importance of correlations in layered nickelates. However, this study specifically focuses on two key issues: (1) the emergence and tunability of a second pocket in the Fermi surface, and (2) the development of electronic charge or spin instabilities. These phenomena can be experimentally manifested by doping or applying external pressure.

The paper is structured as follows: Section (\ref{General_Model}) introduces a comprehensive framework for modeling Nickelate structures with a focus on layered Nickelates. We begin by analyzing the orbital symmetries based on the paramagnetic normal states of well-established, undistorted cubic perovskite Nickelates, and apply these findings to layered Nickelates. We then deduce key orbital degrees of freedom and hybridization effects that can either be neglected or absorbed into the effective tight-binding parameters. This leads us to an effective model for layered Nickelates, which sheds light on the tunability of the Fermi surface. Section (\ref{Lind_Susc}) complements this framework by providing static susceptibility calculations that offer insights into potential ordering phenomena and instabilities within the system.

\section{General Model}\label{General_Model}

Various effective models based on DFT and Wannierization have been developed to capture complex electronic structures of layered Nickelates $RNiO_{2}$. Some of these models utilize up to 17 orbitals to fit the band structure of $RNiO_{2}$, incorporating contributions from Nickel $ 3d $-orbitals, Rare-earth $ 5d $-orbitals, Oxygen $ 2p $-orbitals, and an interstitial $ s $-orbital \cite{Gu2020May,Lechermann2020Feb}. To simplify these models, reductions to four orbitals (such as $ 3d_{x^2-y^2} $, interstitial $ s $, $ 5d_{z^2} $, and $ 5d_{xy} $) \cite{Gu2020May,Lechermann2020Feb}, three orbitals (including $ 3d_{x^2-y^2} $, $ 3d_{z^2} $, and a self-doping orbital) \cite{Kreisel2022Aug,Werner2020Jan,Lechermann2020Oct}, two orbitals ( $ 3d_{x^2-y^2} $ and an interstitial $ s $-orbital) \cite{Chen2023Sep}, or even a single-orbital effective model based on $ 3d_{x^2-y^2} $ \cite{Kitatani2020Aug} have been proposed. However, in this work, we consider the already established orbital symmetry of perovskite Nickelates and deduce the relevant effective orbital responsible for the emergence of an extra Fermi pocket in Layered Nickelates.

In section \ref{sec:RNiOn} we develop generalized modelling framework from multi-orbital description of $RNiO_{n}$, followed by a specific case of layered Nickelates in section \ref{RNiO_2}. Based on the inferences from the relevant symmetries, orbitals and parameters in section \ref{RNiOn}, we propose an effective model capturing trends of the multi-orbital description of layered Nickelates. However, it is to note that the adaptation of this modelling framework can be more general, for example, in Supplementary, where it is qualitatively adapted for 3-Dimensional $RNiO_{3}$.

\subsection{Multi-orbital description of RNiO$_n$} \label{sec:RNiOn}

In this section, we introduce the generalized Hamiltonian formalism for capturing the multi-orbital characteristics of the $RNiO_n$ system, which has $N$ orbitals per site representing atomic orbitals of $R$, $Ni$, and $O$. The non-interacting Hamiltonian, expressed as a block matrix, comprehensively accounts for the electronic structure within a multi-orbital framework.

\begin{figure}[!htbp]
    \centering
    \includegraphics[width=0.5\linewidth]{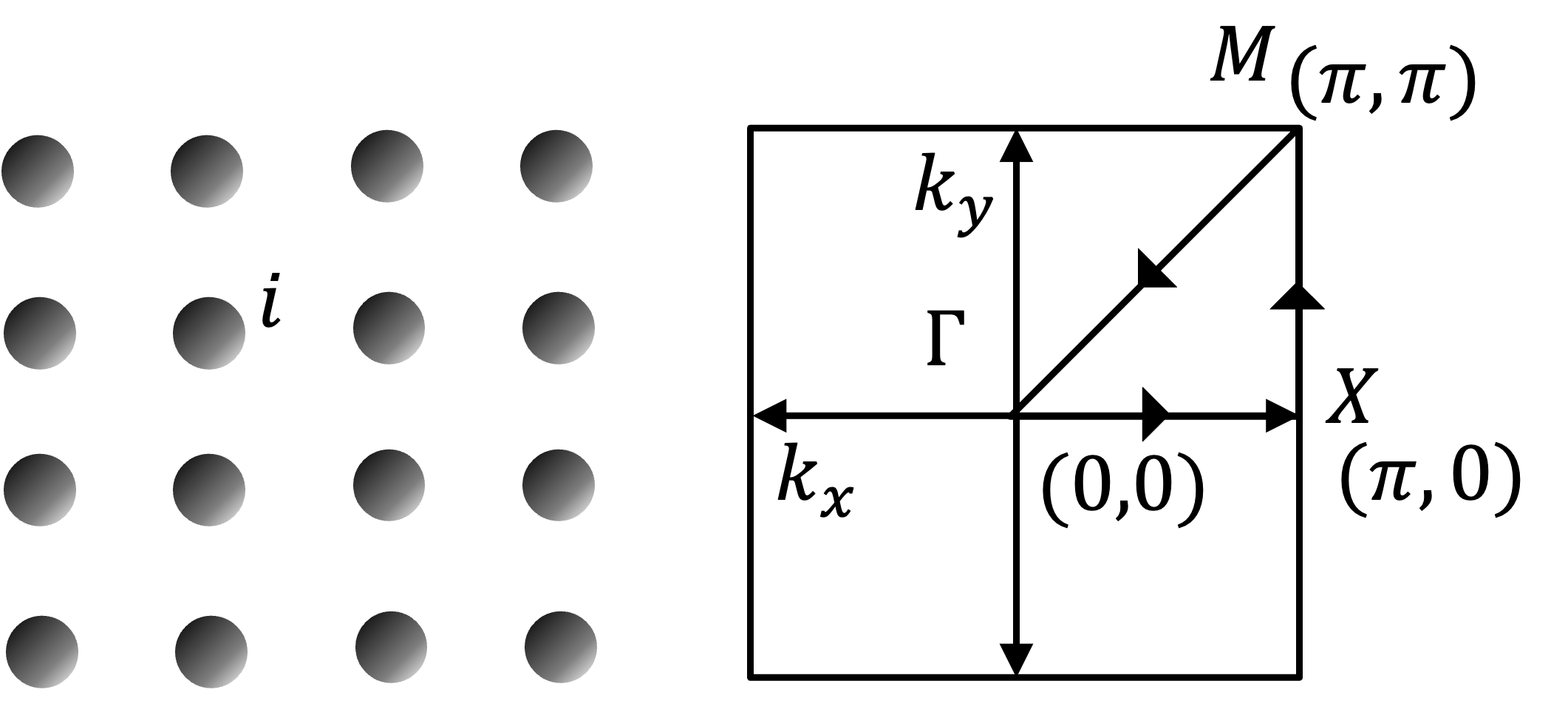}
    \caption{Representative lattice site  $i$ and corresponding Brillouin zone in 2D case. In perovskite Nickelates, lattice site $i$ consists of a Rare Earth and a Nickel atom with 3 Oxygen atoms. In layered Nickelates, lattice site $i$ consists of a Rare Earth and a Nickel atom with 2 Oxygen atoms.}
    \label{fig:Latt_BZ}
\end{figure}

The Hamiltonian $\mathcal{H}$ is given by the following equation:
\begin{equation}\label{eq:General_Model}
\mathcal{H} = \sum_{\sigma} \sum_{i} \Psi_{i}^{\dagger\sigma} \Tilde{\mathcal{E}}_{0} \Psi_{i}^{\sigma} + \sum_{\sigma} \sum_{\langle i j \rangle} \Psi_{i}^{\dagger\sigma} \mathbf{T}_{ij} \Psi_{j}^{\sigma} 
\end{equation}
where the basis set $\Psi_{i}^{\dagger\sigma}$ ($\Psi_{i}^{\sigma}$) is a row (column) matrix with elements being the creation (annihilation) operators $c_{i}^{\dagger\alpha\sigma}$ ($c_{i}^{\alpha\sigma}$) for the orbitals $\alpha \in \{R, Ni, O\}$ and spin $\sigma = \{ \uparrow, \downarrow \}$ at site $i$ (see Fig. \ref{fig:Latt_BZ}). 
Here, $\mathbf{T}_{ij}$ is the hopping matrix between sites $i$ and $j$; $\Tilde{\mathcal{E}}_{0}$ is the on-site energy, which includes diagonal elements $\mathcal{E}_{0}$ and off-diagonal elements $\Delta \mathcal{E}_{0}$, leading to $\Tilde{\mathcal{E}}_{0} = \mathcal{E}_{0} + \Delta \mathcal{E}_{0}$.

Although the three-dimensional (tetragonal) description is crucial for understanding Nickelates, we adopt a two-dimensional (2D) description inspired by Cuprates. This 2D system considers a $NiO_n$ layer coupled to a single rare-earth layer, providing a simplified description that still captures some essential features of the three-dimensional system. This system description can still consider some physical ingredients of a three-dimensional system where the interlayer coupling involves only tunnelling along the direction between Ni and R orbitals (see supplementary for three-dimensional description).

In momentum space, the Hamiltonian in Eq. (\ref{eq:General_Model}) can be rewritten as:
\begin{subequations}\label{eq:Generalized_Hamiltonian_k}
	\begin{align}
		\mathcal{H}  & = \sum_{\sigma} \sum_{\mathbf{k}} \Psi_{\mathbf{k}}^{\dagger\sigma} (\mathcal{E}_{0} + \mathcal{E}_{\mathbf{k}})\Psi_{\mathbf{k}}^{\sigma} \,\,\,\, , \\
		\mathcal{E}_{\mathbf{k}} & = \Delta \mathcal{E}_{0} + \sum_{\mathbf{\delta}} \mathbf{T}_{\delta} e^{i \mathbf{k}\cdot \mathbf{\delta}} ,
	\end{align}
\end{subequations}
where, the summation over $\mathbf{k}$ spans the entire Brillouin zone corresponding to the 2D model. The term $\mathcal{E}_{\mathbf{k}}$ encapsulates both the off-diagonal on-site terms $\Delta \mathcal{E}_{0}$ and the inter-site hopping terms $\mathbf{T}_{\delta} e^{i \mathbf{k}\cdot \mathbf{\delta}}$, where the summation over $\delta$ covers the nearest-neighbor contributions. 

In a more convenient representation, dispersion $\mathcal{E}_{\mathbf{k}}$ is,
\begin{equation}\label{eq:Generalized_Dispersion_k_matrix}
\mathcal{E}_{\mathbf{k}}  = 
\left[\begin{array}{c|c|c}
\Tilde{\varepsilon}^{R}_{\mathbf{k}} & \Tilde{\varepsilon}_{\mathbf{k}}^{RNi} & \Tilde{\varepsilon}_{\mathbf{k}}^{RO} \\
&&\\
\hline
&&\\
\Tilde{\varepsilon}_{\mathbf{k}}^{NiR} & \Tilde{\varepsilon}^{Ni}_{\mathbf{k}} & \Tilde{\varepsilon}_{\mathbf{k}}^{NiO} \\ 
&&\\
\hline
&&\\
\Tilde{\varepsilon}_{\mathbf{k}}^{OR} & \Tilde{\varepsilon}_{\mathbf{k}}^{ONi} & \Tilde{\varepsilon}^{O}_{\mathbf{k}}
    \end{array}\right].
\end{equation}

The dispersion matrix, $\Tilde{\varepsilon}^{AB}_{\mathbf{k}}$, consists of elements that are matrices of size $N_A \times N_B$, where $A$ and $B$ represent atomic species, such as $R$, $Ni$, and $O$. These blocks represent either intra-atomic interactions $\Tilde{\varepsilon}^{A}_{\mathbf{k}}$ or inter-atomic interactions $\Tilde{\varepsilon}^{AB}_{\mathbf{k}}$. The overall dispersion matrix, $\mathcal{E}_{\mathbf{k}}$, is constructed from these blocks to form a larger matrix of size $N \times N$, where $N = N_{R} + N_{Ni} + N_{O}$ represents the total number of degrees of freedom from all the atomic species and hence the total number of orbitals involved.

The dispersion matrix elements are influenced by the symmetry of the orbitals and the overlap integrals. Once the dispersion $\mathcal{E}_{\mathbf{k}}$ is known, various electronic properties can be calculated using Green's function formalism, focusing on the paramagnetic state. The Green's function is defined as $\mathcal{G}_{\mathbf{k}}^{\sigma\sigma'}(\tau) = -\langle \Psi^{\sigma}_{i} (\tau) \Psi^{\dag\sigma'}_{j} (0) \rangle $ and its Fourier transform is given by $\mathcal{G}_{\mathbf{k}}^{\sigma\sigma'}(\omega) = \int d\tau e^{i\omega\tau} \mathcal{G}_{\mathbf{k}}^{\sigma\sigma'}(\tau) $. In paramagnetic case the Green's function is spin diagonal, given by $\mathcal{G}_{\mathbf{k}}^{\sigma\sigma'}(\omega) = \delta_{\sigma\sigma'} \mathcal{G}_{\mathbf{k}}(\omega)$, indicating spin-rotation invariance. Hereafter, we will denote the Green's function as $\mathcal{G}_{\mathbf{k}}(\omega)$, without the spin indices $\sigma \sigma'$ for the sake of clarity. This function is given by $\mathcal{G}_{\mathbf{k}}(\omega) = (\omega - \mathcal{E}_{0}- \mathcal{E}_{\mathbf{k}})^{-1}$, where $\mathcal{E}_{\mathbf{k}}$ is given by Eq. (\ref{eq:Generalized_Dispersion_k_matrix}) and $\mathcal{E}_{0}$ is a constant energy.

We can obtain the spectral function $\mathcal{A}_{\mathbf{k}}(\omega)$ and Density of States (DoS) $\rho(\omega)$ from the imaginary part of the trace of Green's function matrix as,
\begin{subequations}
\begin{align}
    \mathcal{A}_{\mathbf{k}}(\omega) & =\frac{-1}{\pi} Im \,\, Tr(\mathcal{G}_{\mathbf{k}}(\omega + i 0^{+}))  \label{eq:FS} , \\
    \rho(\omega) & = \sum_{\mathbf{k}}  \mathcal{A}_{\mathbf{k}}(\omega) . \label{eq:DoS}
\end{align}
\label{eq:GF}
\end{subequations}
Above Eq. (\ref{eq:FS}) determines the Fermi surface (sharp peaks at the Fermi energy), and Eq. (\ref{eq:DoS}) determines the density of states (DoS).  Thus, by analyzing the spectral function, Fermi surface, and DoS obtained from the Green's function formalism for the generalized Hamiltonian defined by Eq. (\ref{eq:Generalized_Dispersion_k_matrix}), we can gain valuable insights into the electronic structure of $RNiO_{n}$ system.

Our focus is solely on the electronic characteristics of $RNiO_{n}$, setting aside structural complexities (such as orthorhombic distortion, bond deformation, and disproportionation). Further exploiting the octahedral crystal field symmetry where the $e_g$ states are higher in energy than $t_{2g}$ states (implying that $t_{2g}$ states are always filled) and do not play a role near the Fermi surface, we yield relatively minimalistic description applicable to layered Nickelates $RNiO_{2}$ discussed hereafter.

\subsection{4-orbital description of RNiO$_2$}\label{RNiO_2}

In the process of reducing $RNiO_3$ to $RNiO_2$, the apical Oxygen is removed, resulting in the formation of a layered Nickelate phase. The $RNiO_2$ phase has a square planar symmetry with alternating layers of rare-earth $R$ and $NiO_2$ stacked along the crystallographic $c$-axis. The transition from $RNiO_3$ to $RNiO_2$ involves a meta-stable state with a tetragonally distorted octahedral geometry, which influences the electronic structure by lifting the orbital degeneracy in the Nickel $3d$-orbitals. This leads to a specific energy ordering \cite{zuckerman1965crystal} $ \Tilde{\varepsilon}_{3d_{xz}} =\Tilde{\varepsilon}_{3d_{yz}} < \Tilde{\varepsilon}_{3d_{xy}} < \Tilde{\varepsilon}_{3d_{z^2}} < \Tilde{\varepsilon}_{3d_{x^2-y^2}}$ in $RNiO_{2}$. The non-degenerate $\Tilde{\varepsilon}_{3d_{z^2}} < \Tilde{\varepsilon}_{3d_{x^2-y^2}}$ in $RNiO_2$ along with non-magnetic metallic nature suggests $3d^{9}$ ground state. More precisely a composite ground state involves both $3d^{9}$ and $3d^{8}R$ ($R$ is rare-earth) as suggested by experiments such as \textit{O K-edge XAS} \cite{Hepting2020Apr}. Further, controversy about where the doped holes go upon doping  (to $d^{9}\underbar{L}$ or  $d^{8}$) \cite{Nomura2020Oct} leads to the necessity of inclusion of Oxygen orbitals in the proposed model. Since the Nickelates have large $\Delta_{dp} = \Tilde{\varepsilon}_{d}-\Tilde{\varepsilon}_{2p}$ \cite{PhysRevX.10.011024,Hepting2020Apr,PhysRevX.10.011024} compared to Cuprates and perovskites Nickelates, it is plausible that, this property could also be exploited in modelling them.

\begin{figure}[!htbp]
    \centering
    \includegraphics[width=0.5\linewidth]{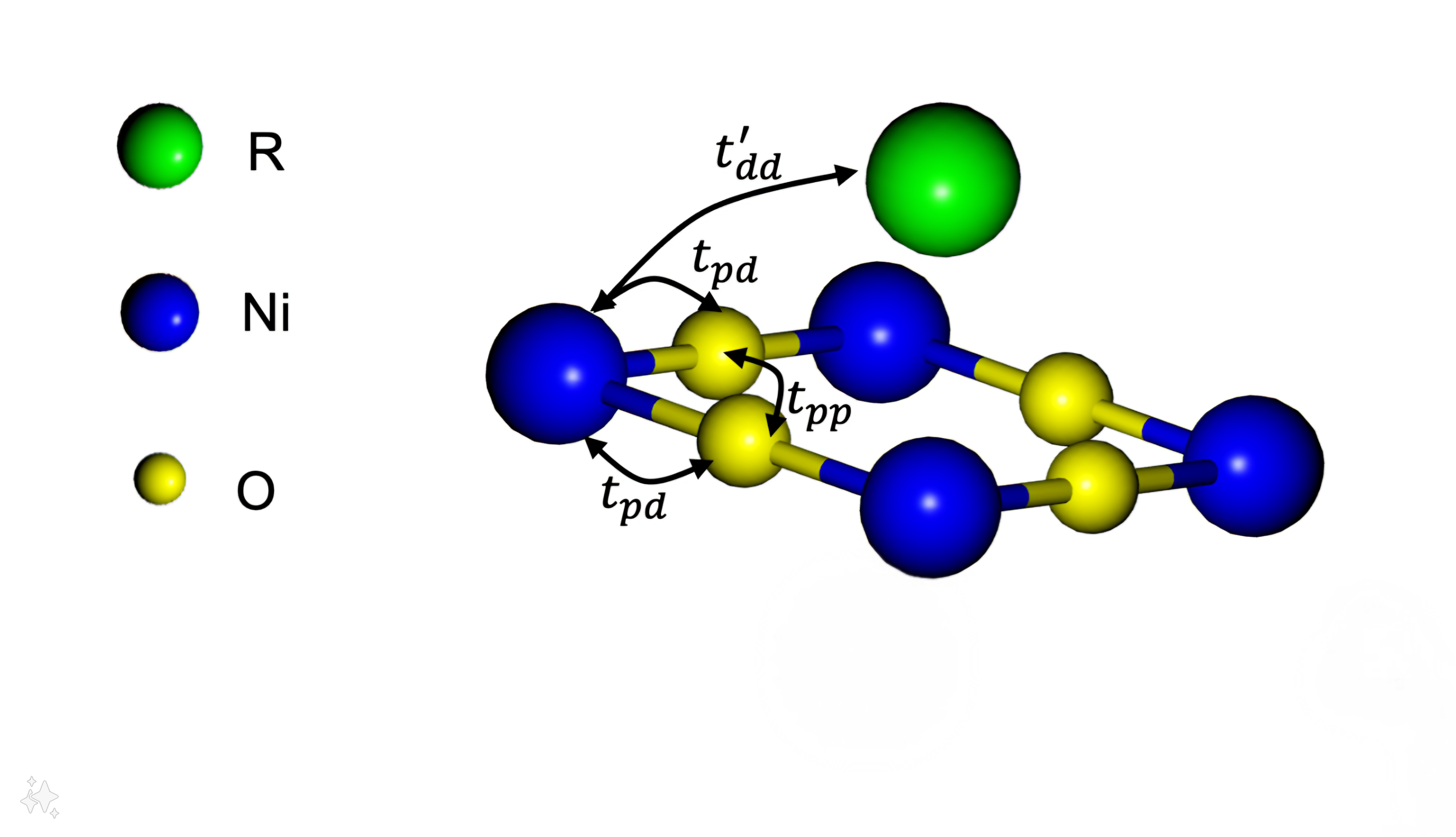}
    \caption{Schematic of in-plane and out-plane hopping parameter in $RNiO_{2}$}
    \label{fig:RNiO_2}
\end{figure}
\begin{figure}[!htbp]
    \centering
    \includegraphics[width=0.5\linewidth]{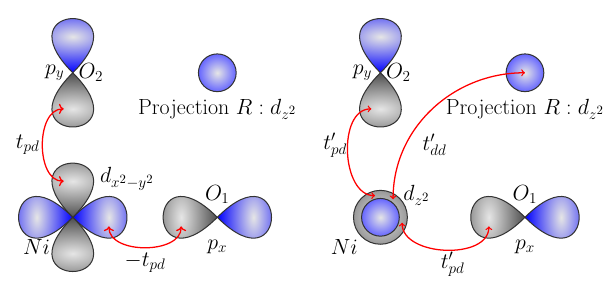}
    \caption{Orbital overlaps and the hopping parameter considering the orbital symmetry of $RNiO_{n}$ and the in-plane Oxygen atoms $O_{1}$ and $O_{2}$.} 
    \label{fig:Orbital_Overlaps_RNiO2}
\end{figure}

Thus, as noted earlier in layered Nickelates $3d_{x^2-y^2}$ and $3d_{z^2}$ lie near Fermi-level similar to perovskite Nickelates  \cite{Dhaka2015Jul} and Cuprates. Further, both parent perovskite and layered Nickelates have approximately similar $e_g$ bandwidth of $3 \,eV$ \cite{Georgescu2019Jul, PhysRevB.85.245131, Liu2013Nov, PhysRevLett.95.127204, PhysRevX.11.011050}. In perovskite nickelates, the bandwidth decreases with confinement, leading to an increased charge transfer gap $\Delta_{d-p}$. This is associated with the shrinking ionic radius of $R$ due to Lanthanide Contraction with increasing atomic number and an expanding lattice mismatch on which the layer is grown \cite{Georgescu2019Jul, PhysRevB.85.245131, Liu2013Nov, PhysRevLett.95.127204}. Conversely, in $RNiO_2$, the bandwidth consistently expands across the Lanthanide series. Thus, such rare-earth doping is attributed to a reduction in both in-plane \cite{PhysRevB.102.205130} and inter-plane \cite{PhysRevX.11.011050} lattice constants in layered Nickelates, leading to increased orbital overlap, hybridization, and overall bandwidth. A recent development in the debate about questioning the necessity of $3d_{z^2}$ orbital to describe superconductivity focuses primarily on the $3d_{x^2-y^2}$ orbital \cite{Si2024Aug} supported by ARPES measurement. However, the energetically high-spin scenario is more favourable than the low-spin scenario when transitioning from $d^{9} \rightarrow d^{8}$\cite{Plienbumrung2021Mar}. This implies that out of two channels (in-plane within $NiO_{2}$ layer or inter-plane through Rare-earth layer), there could be competition which leads to the emergence of a second pocket \cite{Zeng2020Oct, Zeng2022Feb, Osada2020Dec, Osada2021Nov, Li2019Aug, Li2020Jul}.

\begin{figure*}[!htbp]
\centering
   \includegraphics[width=1\linewidth]{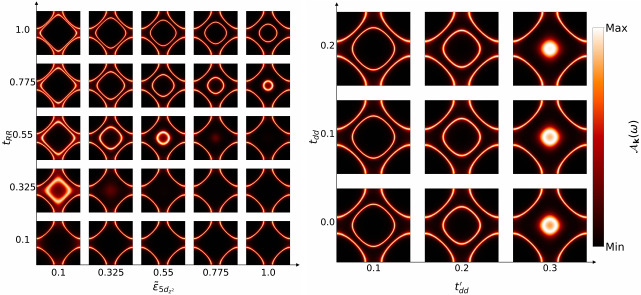}
    \caption{Fermi surface evolution  when $\Tilde{\varepsilon}_{3d_{z^2}}=-1.5\,eV$ and $ \Tilde{\varepsilon}_{3d_{x^2-y^2}} = -1.0\,eV$ in $RNiO_{2}$. Left Panel: Varying $t_{RR}$ and $\Tilde{\varepsilon}_{5d_{z^2}}$ for a  fixed $t^{\prime}_{dd} = 0.2\,eV$. Right panel: Varying $t^{\prime}_{dd}$ and $t_{dd}$ for a fixed $t_{RR} = 0.775 \,eV$ and $\Tilde{\varepsilon}_{5d_{z^2}} = 0.55 \,eV$. Each Fermi-surface extend from $-\pi$ to $\pi$ in $(k_x,k_y)$ plane.}
    \label{fig:FS_Evol_1.5}
\end{figure*}
To initiate examination of this competition and tunability, we draw analogies from orbital symmetry of related pervoskite Nickelates and consider $5d_{z^2}$, $3d_{x^2-y^2}$, $3d_{z^2}$, $2p_{x}$ and $2p_{y}$ orbitals. The 2D orbital overlap of these orbitals is as depicted in  Fig. \ref{fig:Orbital_Overlaps_RNiO2}. Therefore, using the Hamiltonian Eq. (\ref{eq:Generalized_Hamiltonian_k}) we get blocks in Eq. (\ref{eq:Generalized_Dispersion_k_matrix}) as,
    \begin{subequations}\label{Eq:GeneralizedDispersion_RNiO_2}
    \begin{align}
        \Tilde{\varepsilon}^{O}_{\mathbf{k}}  = \Tilde{\varepsilon}_{2p} 
        +\left[\begin{array}{cc}
            0 &  t_{pp} s_{k_x} s^{*}_{k_y}\\
            t_{pp} s^{*}_{k_x} s_{k_y}  & 0 \end{array}\right] 
            \label{Eq:GeneralizedDispersion_RNiO_2_O} \,\,\,\, , \\
       \Tilde{\varepsilon}^{NiO}_{\mathbf{k}}  = \left[\begin{array}{cc}
            -t^{\prime}_{pd}s_{k_x} & -t^{\prime}_{pd}s_{k_y} \\
            t_{pd} s_{k_x} & -t_{pd} s_{k_y}  \end{array}\right]  
            \label{Eq:GeneralizedDispersion_RNiO_2_Ni_O} \,\,\,\, ,\\
      \Tilde{\varepsilon}^{Ni}_{\mathbf{k}}  =  \left[\begin{array}{cc}
            \Tilde{\varepsilon}_{3d_{z^2}} & 0 \\
            0 & \Tilde{\varepsilon}_{3d_{x^2-y^2}} \\ \end{array}\right] 
        + \left[\begin{array}{cc}
            0 &  t_{dd}\\
            t_{dd} & 0 \\ \end{array}\right]
            \label{Eq:GeneralizedDispersion_RNiO_2_Ni} \,\,\,\, ,\\
        \Tilde{\varepsilon}^{R}_{\mathbf{k}}  = \Tilde{\varepsilon}_{5d_{z^2}} + t_{RR}(\cos(k_x)+\cos(k_y))\,\,\,\, , \\
        \Tilde{\varepsilon}^{RNi}_{\mathbf{k}}  = \left[\begin{array}{cc}
            t^{\prime}_{dd}s_{k_x}s_{k_y}  & t^{\prime\prime}_{dd}s_{k_x}s_{k_y}\end{array}\right] \,\,\,\, ,\label{Eq:GeneralizedDispersion_RNiO_2_R_Ni}
    \end{align}
\end{subequations}
where 
\begin{equation}
 s_{k_{\nu}} =(1-e^{i k_{\nu}})
\end{equation}
and its complex conjugate $s_{k_{\nu}}^* =(1-e^{-i k_{\nu}})$.
\begin{figure*}[!htbp]
    \centering
    \includegraphics[width=1\linewidth]{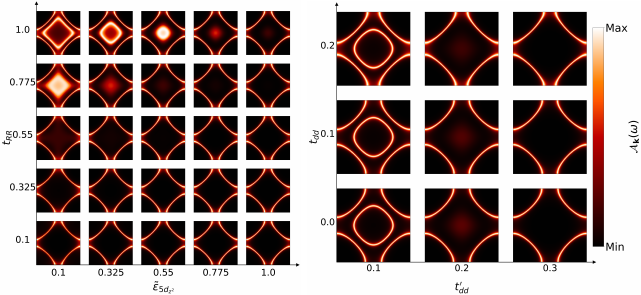}
    \caption{Fermi surface evolution when $\Tilde{\varepsilon}_{3d_{z^2}}=-0.5\,eV$ and $ \Tilde{\varepsilon}_{3d_{x^2-y^2}} = -1.0\,eV$ in $RNiO_{2}$. Left Panel: Varying $t_{RR}$ and $\Tilde{\varepsilon}_{5d_{z^2}}$ for a  fixed $t^{\prime}_{dd} = 0.2\,eV$. Right panel: Varying $t^{\prime}_{dd}$ and $t_{dd}$ for a fixed $t_{RR} = 0.775 \,eV$ and $\Tilde{\varepsilon}_{5d_{z^2}} = 0.55 \,eV$. Each Fermi-surface extend from $-\pi$ to $\pi$ in $(k_x,k_y)$ plane.}
    \label{fig:FS_Evol_0.5}
\end{figure*}

As shown in Figs. \ref{fig:RNiO_2} and \ref{fig:Orbital_Overlaps_RNiO2}, the parameter labeled as $t_{pd}$ represents the hopping between $3d_{x^2-y^2}$ and $2p_{x/y}$ in the $x-y$ plane. $t^{\prime}_{pd}$ denotes the hopping associated with $3d_{z^2}$ and $2p_{x/y}$. The parameters $t^{\prime}_{dd}$ and $t^{\prime\prime}_{dd}$ refer to the hopping involving $5d_{z^2}-3d_{z^2}$ and $5d_{z^2}-3d_{x^2-y^2}$, respectively. The $t_{dd}$ describes the hybridization connecting $3d_{z^2}$ and $3d_{x^2-y^2}$ (not shown in Fig. \ref{fig:RNiO_2}). Finally, $t_{pp}$ represents the hopping interaction of $2p_{x}$ and $2p_{y}$ in the $x-y$ plane, while $t_{RR}$ corresponds to the hopping within the Rare-earth's $5d_{z^2}$. The overlap values for the $RO$ orbitals are set to zero in the dispersion term based on symmetry considerations, as illustrated in Fig. \ref{fig:Orbital_Overlaps_RNiO2}, which indicate cancellation of overlap for these orbitals (due to bonding and anti-bonding portions of $3d_{x^2-y^2}$ overlapping with the symmetric projection of Rare-earth $5d_{z^2}$). Based on the typical parameters reported from various ab initio calculations as listed in \ref{App:Parameters}, common parameters used in the calculation are tabulated in Table. \ref{tab:used_para}. 
\begin{table}[!htbp]
\centering
\begin{tabular}{|c|c|c|c|c|c|c|}
\hline
$t_{pd}$ & $t_{pp}$ & $t_{dd}$ & $t^{\prime}_{dd}$ & $t_{RR}$ & $\Tilde{\varepsilon}_{3d_{x^2-y^2}}-\Tilde{\varepsilon}_{2p} $  & $\Tilde{\varepsilon}_{5d_{z^2}} $\\ \hline
$1.2\,eV$   & $0.6\,eV$   & $0.0$-$0.2\,eV$   & $0.1$-$0.3\,eV$ & $0.1$-$1.0\,eV$ & $6\,eV$  &   $0.1$-$1.0\,eV$                                                     \\ \hline
\end{tabular}
\caption{Summary of common model parameters used (See also \ref{App:Parameters} for a non-exhaustive list of some reported parameters from various \textit{ab-initio} calculations).}
\label{tab:used_para}
\end{table}
Here, we set an energy difference of 6 \,eV between the $3d_{x^2-y^2}$ and $p$ orbitals, establishing the typical scale for electronic level degeneracy and the positioning of the low-lying Oxygen level. Consequently, the reference $Ni:3d$ occupancy for $n_{3d_{x^2-y^2}} + n_{3d_{z^2}}$ is approximately 3.2 electrons.

Because of the extended nature of the Rare-earth orbital, the hopping terms $t^{\prime}_{pd}$ and $t^{\prime\prime}_{dd}$ are small and thus set to zero. Additionally, the Oxygen $2p$ levels ($\tilde{\varepsilon}_{2p}$) lie significantly below the Fermi level $\tilde{\varepsilon}_F$. These and other parameter values are provided in  \ref{App:Parameters} based on various reported \textit{ab-initio} calculations. Due to the extended nature of the Rare-earth $5d$ orbitals, we explicitly include the direct $R-R$ hopping, denoted as $t_{RR}$, which can be tuned alongside $t^{\prime}_{dd}$ and $t_{dd}$ to mimic the effect of reduction of apical Oxygen and compression along the $z$-axis. At fixed hybridization between $3d_{z^2}-3d_{x^2-y^2}$ and   $3d_{z^2}-5d_{z^2}$, tuning relative position of Rare-earth $5d_{z^2}$ along with $t_{RR}$  and vice-versa helps to understand the interdependence of the $3d_{z^2}$ orbital and the rare-earth $5d_{z^2}$ orbital. More specifically, to understand the variation in input parameters quantitatively in layered Nickelates, such as in $PrNiO_{2}$, we can refer to Table 1 of the reference \cite{DiCataldo2024May}, where there is explicit mention of changes in lattice and hopping parameters with respect to applied pressure. In the extreme case of $150$ GPa pressure, the onsite energies and in-plane hopping parameters increase by up to $60 \% $. Pressure along the $z$-axis results in $\tilde{\varepsilon}_{3d_{z^2}} > \tilde{\varepsilon}_{3d_{x^2-y^2}}$ below the Fermi level. This conditions tends to reduce the in-plane hopping . Hence Fermi-surface variation with $t^{'}_{dd}$ and $t_{dd}$ is dormant (see Fig.\ref{fig:FS_Evol_0.5}) while in lowered $\tilde{\varepsilon}_{3d_{z^2}} < \tilde{\varepsilon}_{3d_{x^2-y^2}}$ condition, the $3d_{z^2}$ acts as a drain for the in-plane conducting electron and contributes as a mediator for super-exchange from $Ni-Ni$ across the layer.  It is hence more robust with the change of $t^{'}_{dd}$ and $t_{dd}$ (see Fig.\ref{fig:FS_Evol_1.5}). A similar trend is also reported in $327$ layered Nickelates ($La_{3}Ni_{2}O_{7})$ at $30$ GPa \cite{Fan2024Jul} where there are evidence of decoupled $3d_{z^2}$ under pressure considering full-band initialization for DFT-based calculation. This model is exemplary in understanding the tunability of the Fermi surface corresponding to the variation in electronic structure due to external pressure or doping. It is to be noted that the emergent secondary pocket in the Fermi surface in this model exists due to interaction between $3d_{z^2}$ and $5d_{z^2}$ while the primary pocket is mainly due to $3d_{x^2-y^2}$. Further, by comparing Fermi-surface variation with variation in  $\tilde{\varepsilon}_{5d_{z^2}}$ and $t_{RR}$ in Figs. \ref{fig:FS_Evol_1.5}, and \ref{fig:FS_Evol_0.5}, we can observe that the two-pocket Fermi-surface is more sensitive to the relative difference between $\tilde{\varepsilon}_{3d_{x^2-y^2}}$ and $\tilde{\varepsilon}_{3d_{z^2}}$ due to the hybridization between $5d_{z^2}$ and $3d_{z^2}$. Consequently, in the case when $\tilde{\varepsilon}_{3d_{z^2}} < \tilde{\varepsilon}_{3d_{x^2-y^2}}$ the emergence of second Fermi-pocket is more favourable. Hence, this analysis can infer $3d_{z^2}$ as an anchoring orbital level. The relative position of $3d_{z^2}$ to $3d_{x^2-y^2}$ can decide the role of $5d_{z^2}$ in the emergence of an additional Fermi-pocket in the Nickelates system.

\subsection{Effective 2-orbital description of $RNiO_{2}$}\label{RNiOn}

\begin{figure}[!htbp]
    \centering
    \includegraphics[width=0.8\linewidth]{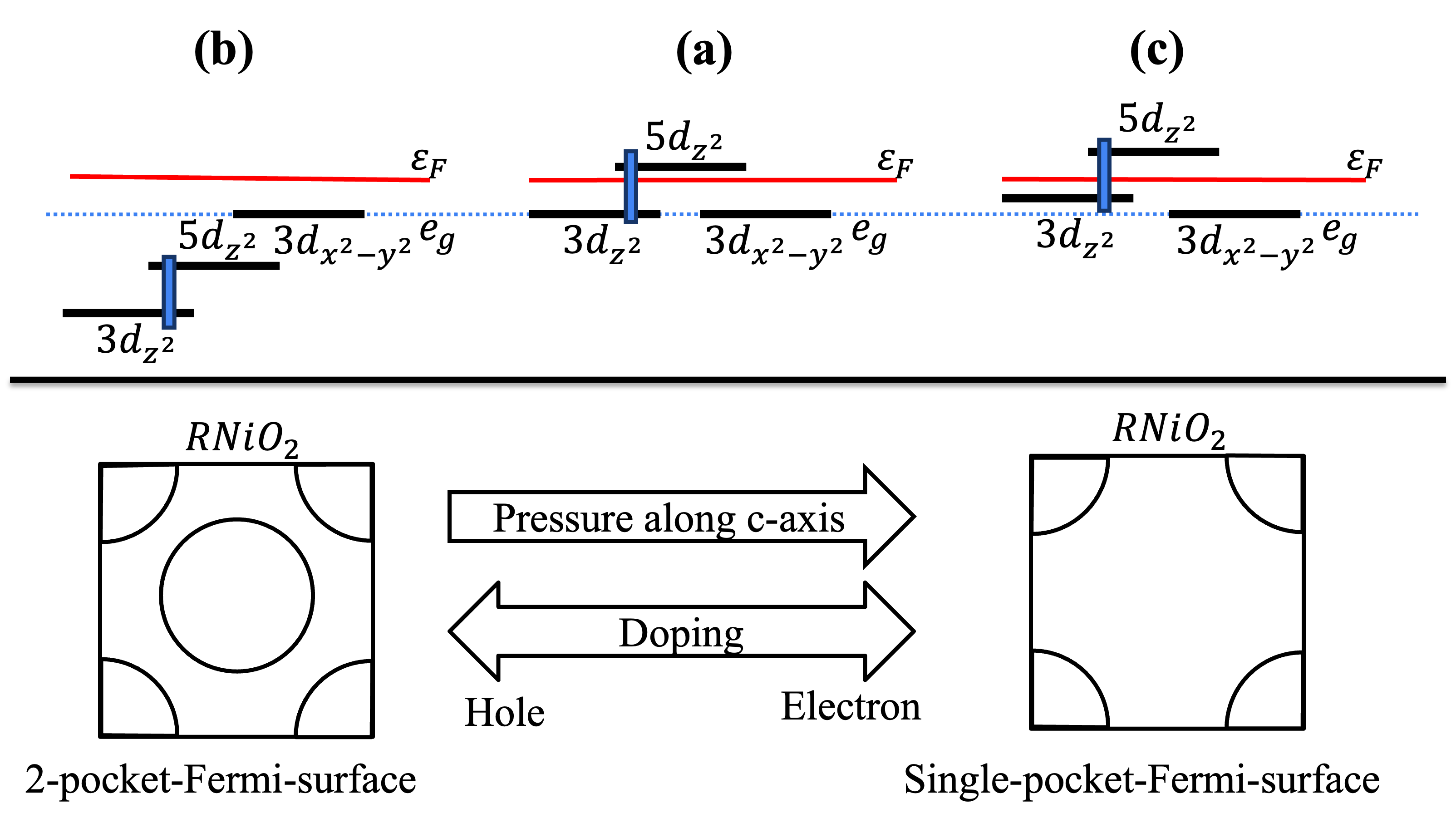}
    \caption{Anchoring effect of Ni:$3d_{z^{2}}$ with R:$5d_{z^2}$  with condition \textbf{(a)} $\Tilde{\varepsilon}_{3d_{z^2}}=\Tilde{\varepsilon}_{3d_{x^2-y^2}}$,  \textbf{(b)} $\Tilde{\varepsilon}_{3d_{z^2}}<\Tilde{\varepsilon}_{3d_{x^2-y^2}}$,  \textbf{(c)} $\Tilde{\varepsilon}_{3d_{z^2}}>\Tilde{\varepsilon}_{3d_{x^2-y^2}}$. Mimicking the external doping and applied pressure.}
    \label{fig:RNiOn_results}
\end{figure}

In the previous sections, we have shown the anchoring effect of $3d_{z^2}$ and its resemblance to the impact of doping and pressure on the evolution of the Fermi surface. As illustrated in Fig. \ref{fig:RNiOn_results}, depending on the change in magnitude of crystal field splitting due to pressure as well as the type of doping (hole or electron), Nickel's $3d_{z^2}$ can either anchor down or up the Rare-earth's $5d_{z^2}$ in terms of energy, leading to the emergence of an additional Fermi pocket.

\begin{figure}[!htbp]
    \centering
    \includegraphics[width=0.6\linewidth]{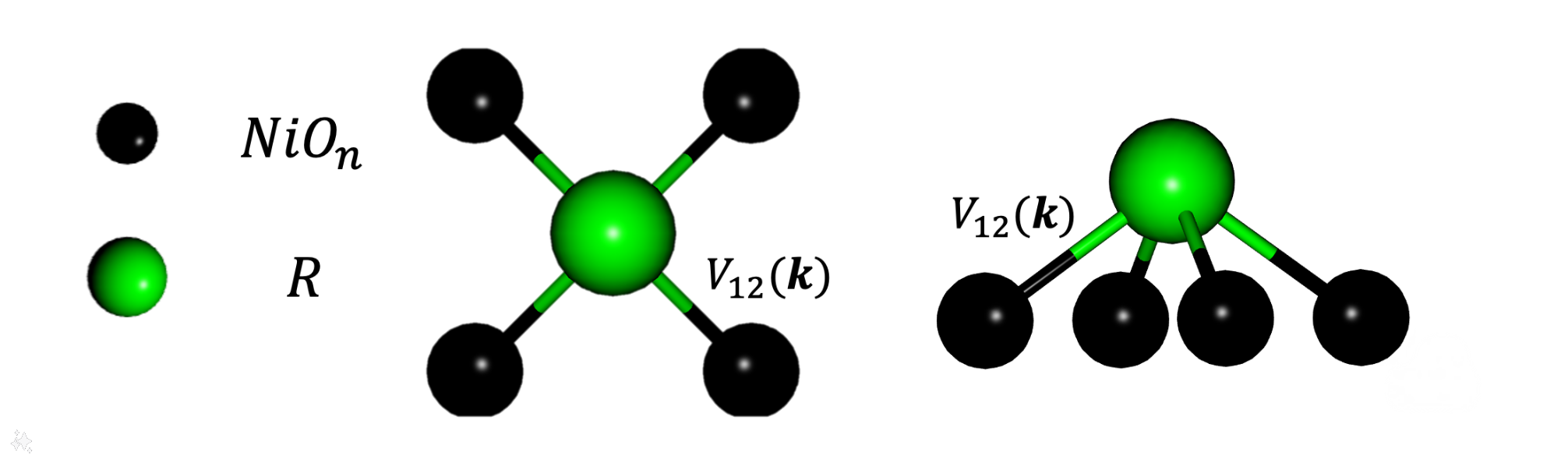}
    \caption{Unit cell of effective $RNiO_{2}$ system.}
    \label{fig:RNiOn}
\end{figure}

One may now ask, what could be an effective model that mimics the Fermi surface's emergent features and trends, as discussed in the section \ref{RNiO_2} for $RNiO_{2}$? We propose that the adequate multi-orbital description of $RNiO_{n}$ along with the consideration of the presence of a second pocket Fermi Surface, can be simplified considering a system consisting of two 2D square lattices (one equivalent to the Rare-Earth layer and another equivalent $NiO_{n}$ layer with pseudo-orbital equivalent to $3d_{x^2-y^2}+2p_{x/y}$) as shown in schematic Fig.\ref{fig:RNiOn}.

Theoretically, this sort of reduction can be done using the downfolding method (see \ref{App:Downfolding} for more details) of full Hamiltonian Eq. (\ref{eq:Generalized_Dispersion_k_matrix}) to effective Hamiltonian of form, 
\begin{equation}\label{eq:two_band_dispersion_RNiO_2}
\mathcal{E}_{eff}(\mathbf{k}) = \left[\begin{array}{c|c}
\mathcal{\Tilde{E}}_{1} (\mathbf{k}) & V_{12}(\mathbf{k}) \\
&\\
\hline
&\\
V_{12}^{*}(\mathbf{k}) & \mathcal{\Tilde{E}}_{2} (\mathbf{k})
\end{array}\right]
\end{equation}
such that,
\begin{subequations}\label{eq:explicit_self_energies_RNiO_2}
    \begin{align}
        \mathcal{\Tilde{E}}_{1} (\mathbf{k})  & = \Tilde{\varepsilon}^{R}_{\mathbf{k}} + \Tilde{\varepsilon}_{\mathbf{k}}^{RO}  [\Tilde{\varepsilon}^{O}_{\mathbf{k}}]^{-1} \Tilde{\varepsilon}_{\mathbf{k}}^{OR} \,\,\,\, , \\
        \mathcal{\Tilde{E}}_{2} (\mathbf{k})  & = \Tilde{\varepsilon}^{Ni}_{\mathbf{k}} + \Tilde{\varepsilon}_{\mathbf{k}}^{NiO}  [ \Tilde{\varepsilon}^{O}_{\mathbf{k}}]^{-1} \Tilde{\varepsilon}_{\mathbf{k}}^{ONi} \,\,\,\, , \\
       V_{12}(\mathbf{k}) & = \Tilde{\varepsilon}_{\mathbf{k}}^{RNi}+  \Tilde{\varepsilon}_{\mathbf{k}}^{RO}  [ \Tilde{\varepsilon}^{O}_{\mathbf{k}}]^{-1} \Tilde{\varepsilon}_{\mathbf{k}}^{ONi} \,\,\,\, .
    \end{align}
\end{subequations} 
This reduction to effective Hamiltonian is restricted to the low-energy description of the system and is allowed only when the energy of ligand $(O)$ is significantly lower than the Fermi level related to other atoms $(R , Ni)$ in the compound. This also assumes that the valence fluctuation on Oxygen can be relatively small, which is the usual case in most layered Nickelates. 

For a specific case of $RNiO_{2}$, we substitute explicit terms from Eq. (\ref{Eq:GeneralizedDispersion_RNiO_2}) in Eq. \ref{eq:explicit_self_energies_RNiO_2} along with the assumptions as mentioned in section \ref{RNiO_2} that due to the influence of extended Rare-earth orbit, $t^{\prime}_{pd}$ and $t^{\prime\prime}_{dd}$ are small and hence are set to zero. Further, as there is no overlap between Rare-earth and Oxygen, hence we can set $\Tilde{\varepsilon}^{RO}_{\mathbf{k}}=0$. We thus obtain a generalized form of dispersion as, 
\begin{equation}\label{eq:R_R_eq}
\mathcal{\Tilde{E}}_{1} (\mathbf{k}) = \Tilde{\varepsilon}_{5d_{z^2}} + t_{RR}*( \cos(k_x) + \cos(k_y)) \,\,\,\, ,
\end{equation} 
corresponding to Rare-Earth,
\begin{equation}\label{eq:Ni_Ni_eq}
\mathcal{\Tilde{E}}_{2} (\mathbf{k}) = t_{dd} + \frac{2t_{pd}^2}{\Tilde{\varepsilon}_{2p}}(\cos(k_x) + \cos(k_y))) \,\,\,\, ,
\end{equation}
corresponding equivalent $NiO_{n}$ layer and 
\begin{equation}\label{eq:R_Ni_eq}
    V_{12}(\mathbf{k}) = t^{\prime}_{dd}*(1-e^{k_{x}})*(1-e^{k_{y}})
\end{equation}   
is interlayer hybridization.

Therefore, the proposed effective model leads to an effective two-orbital model with dispersion given by equation (\ref{eq:two_band_dispersion_RNiO_2}). This model is achieved through an effective two-layer 2D square lattice system with interlayer hybridization $V_{12}(\mathbf{k})$. The key component of this approach is to incorporate the effective in-plane $Ni-O-Ni$ hopping term Eq. (\ref{eq:Ni_Ni_eq}), which in principle is shared by both families of perovskite and layered Nickelates due to the integration of Oxygen's contribution to the effective $NiO_{2}$ layer. Additionally, the formulation explicitly considers the direct $R-R$ hopping term Eq. (\ref{eq:R_R_eq}), which is influenced by the presence of a dopant, which can enhance or reduce the interlayer hybridization.

\begin{figure}[!htbp]
    \centering
    \includegraphics[width=0.8\linewidth]{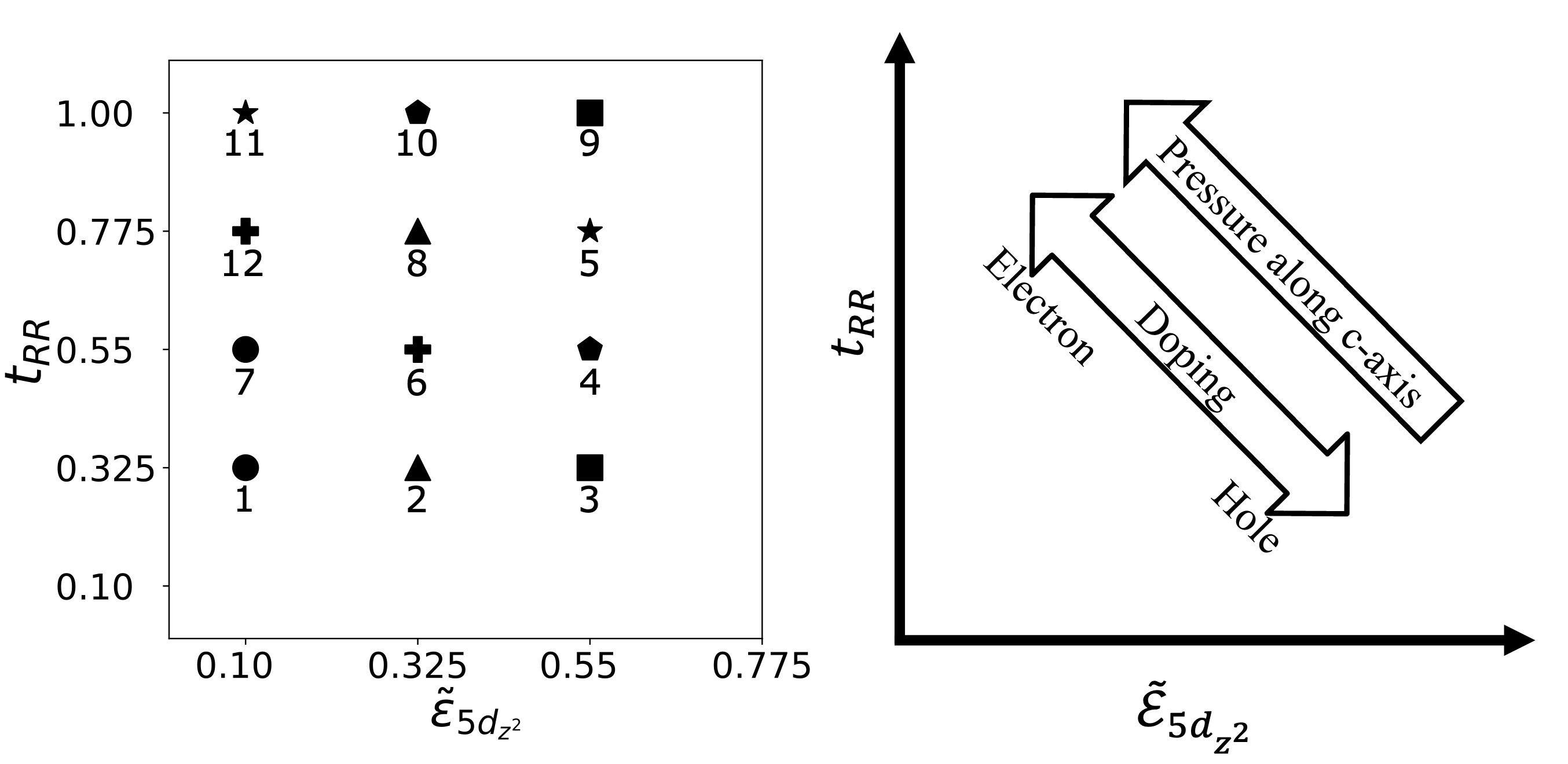}
    \caption{Tuning parameters (Left panel) and schematic of its trend (Right panel) under pressure and doping (see also Table. \ref{tab:trend}).}
    \label{fig:Trend_OT}
\end{figure}

\begin{figure*}[!htbp]
    \centering
    \includegraphics[width=1\linewidth]{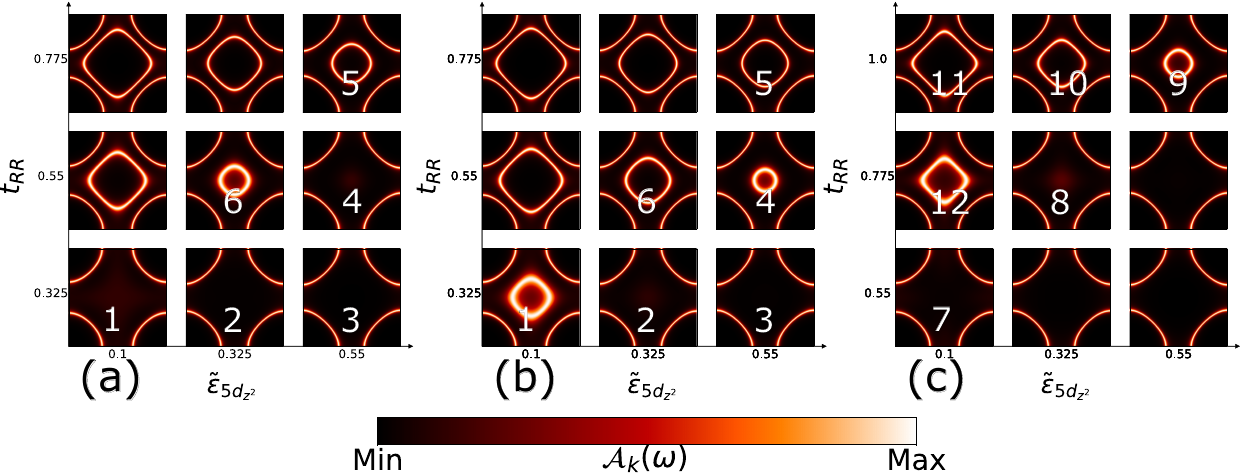}
    \caption{Fermi surface evolution for $RNiO_2$ from our proposed effective model given by dispersion Eq. (\ref{eq:two_band_dispersion_RNiO_2}) capturing the similar features of Fermi-surface evolution of multi-orbital model proposed in section (\ref{RNiO_2}) for $RNiO_{2}$ with condition \textbf{(a)} $\Tilde{\varepsilon}_{3d_{z^2}}=\Tilde{\varepsilon}_{3d_{x^2-y^2}}$,  \textbf{(b)} $\Tilde{\varepsilon}_{3d_{z^2}}<\Tilde{\varepsilon}_{3d_{x^2-y^2}}$,  \textbf{(c)} $\Tilde{\varepsilon}_{3d_{z^2}}>\Tilde{\varepsilon}_{3d_{x^2-y^2}}$ mimicking the doping and pressure effect (see Fig. \ref{fig:RNiOn_results}). The number in Figures refers to the specific choice of the parameter as mentioned in Fig. \ref{fig:Trend_OT} (will also be used in section $\ref{Lind_Susc}$ for Lindhard susceptibility calculation see Fig. \ref{fig:Chi_Evol}).}
    \label{fig:Eq_FS_Evol}
\end{figure*}  

\begin{figure*}[!htbp]
    \centering
    \includegraphics[width=1\linewidth]{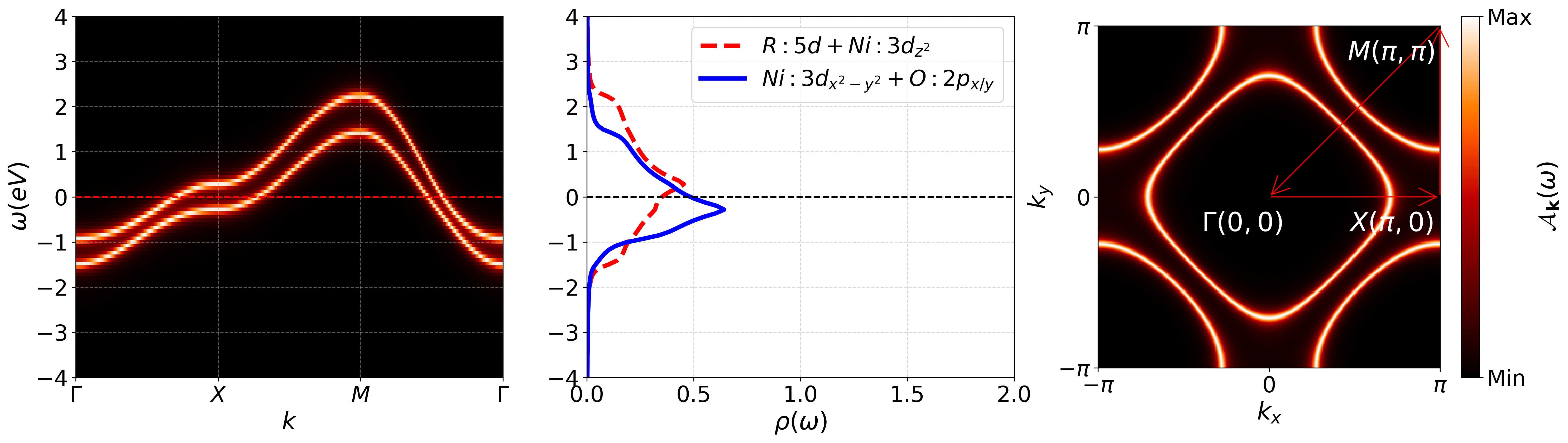}
    \caption{Bands Structure, Orbital-Resolved Density of States and Fermi-Surface from effective 2-orbital description with dispersion given by Eq. (\ref{eq:two_band_dispersion_RNiO_2})  using parameters from ref. \cite{Botana2020Feb}}
    \label{fig:Eq_Bands_DoS_FoS}
\end{figure*}

To check the equivalence of the effective model, we use a downfolded dispersion matrix Eq. (\ref{eq:two_band_dispersion_RNiO_2}) and calculate Fermi-surface evolution using the spectral function from the Green's function formalism Eq. (\ref{eq:GF}) as shown in Fig. \ref{fig:Eq_FS_Evol}. We consider three distinct cases for the onsite energies $ \Tilde{\varepsilon}_{3d_{z^2}} $ and $ \Tilde{\varepsilon}_{3d_{x^2-y^2}} $, depicted in Fig. \ref{fig:Eq_FS_Evol}: $1.)\,\,  \Tilde{\varepsilon}_{3d_{z^2}} = \Tilde{\varepsilon}_{3d_{x^2-y^2}} $ (Left panel): This occurs in an undistorted octahedral field environment,  and is typically caused by Oxygen ligands, which maintain symmetry and equalizes the $ 3d $-orbital energies. $2.)\,\,  \Tilde{\varepsilon}_{3d_{z^2}} < \Tilde{\varepsilon}_{3d_{x^2-y^2}} $ (Middle panel): square planar field symmetry, lowering the energy of the $ 3d_{z^2} $ orbital relative to the $ 3d_{x^2-y^2} $ orbital. $3.)\,\,  \Tilde{\varepsilon}_{3d_{z^2}} > \Tilde{\varepsilon}_{3d_{x^2-y^2}} $ (Right panel): Under high pressure, orbital overlap increases, raising the energy of $ 3d_{z^2} $, which leads to an increase in bandwidth and a higher $ 3d_{z^2} $ energy. These cases mimic the effects of dopants (like Oxygen) and applied pressure (see Fig. \ref{fig:RNiOn_results}). Also, using parameters from ref. \cite{Botana2020Feb} we calculate the band-structure, density of state, and Fermi-surface as plotted in Fig. \ref{fig:Eq_Bands_DoS_FoS}. By comparison of Fig. \ref{fig:Eq_FS_Evol} with Figs. \ref{fig:FS_Evol_1.5}, and \ref{fig:FS_Evol_0.5}, we can observe that both multi-orbital model, as well as effective two-orbital models, have a similar emergence of two-pocket Fermi-surface and Fermi-Surface evolution, as discussed in sections (\ref{RNiO_2}) and (\ref{RNiO_2}). This can be well understood by this effective two-band model as mentioned in Eq. (\ref{eq:two_band_dispersion_RNiO_2}). The layer corresponding to the equivalent square lattice of $NiO_{2}$ has the band-contribution from $3d_{x^2-y^2}$ and $2p_{x/y}$ orbitals. The other layer corresponding to the equivalent square lattice of $R-Ni-O$ has band-contribution from $3d_{z^2}$ and $5d_{z^2}$ orbitals, with inter-layer hybridization $V_{12}$, which can be momentum dependent due to interstitial position of Rare-Earth. We also note that the two Van-Hove singularities (two peaks in the density of states (DOS) in Fig. \ref{fig:Eq_Bands_DoS_FoS}) have the dominant contribution of $R-Ni$ and $Ni-O$ orbitals. While tuning the parameters in these systems, the Fermi surface changes. The appearance of a secondary Fermi pocket is realized when the Van-hove singularity of $R-Ni$ moves much closer to the Van-hove singularity of $Ni-O$. Indeed, the downfolding method will lose the occupancy count of the band under the scrutiny of empirical evidence of input parameters and full-band Density Functional Theory \textit{(ab-initio)} analysis. However, it will preserves a consistent trend of inter-dependence of parameters and the order of magnitude and effectively gives the emergence and evolution of a two-pocket-Fermi surface

Thus, we have examined a generalized phenomenological model based on an understanding of the relevant multi-orbital model, which consists of Nickel $3d_{z^2}$ and Rare-earth $5d_{z^2}$ in layered Nickelates (see Fig. \ref{fig:RNiOn_results}). This has led to an effective two-orbital model for $RNiO_{2}$ . Our phenomenological understanding emphasizes the role of $3d_{z^2}$ as an anchoring energy level, which in our model can be explicitly captured in the interlayer hybridization term $V_{12}(\mathbf{k})$ (see Fig. \ref{fig:RNiOn}) in dispersion equation (\ref{eq:two_band_dispersion_RNiO_2}). This effective model retains the Fermi-surface evolution features of the multi-orbital model, even with the qualitative tuning of doping and pressure. Therefore, it can be utilized to calculate other physical observables. For example, in the subsequent section (\ref{Lind_Susc}), we use it to calculate the static susceptibility in the considered system, helping us understand the nature of electronic excitation and instabilities.

\section{Spin and Charge susceptibilities}\label{Lind_Susc}
In this section, we analyze the spin and charge channels of instabilities by evaluating the Lindhard susceptibility within the framework of our effective two-orbital model Eq. (\ref{eq:two_band_dispersion_RNiO_2}). We aim to determine how well this two-orbital model captures the key physical phenomena, such as spin and charge instabilities relevant to high-$T_c$ superconductivity \cite{Smith1983Sep}. These instabilities often emerge from the complex interplay between charge, spin, and orbital degrees of freedom. They can lead to transitions in the pairing symmetry and gap structure of the superconducting state. In particular, spin and charge fluctuations may drive the system towards different ordered phases, such as charge-density waves, spin-density waves, or superconducting phases. Observables like susceptibility can provide valuable insights into these instabilities by probing the behavior in charge or spin channels. 

We define,  multiorbital site-dependant charge susceptibility as,
\begin{equation}\label{eq:charge_sus_tau}
    \chi^{AB}_{ij}(\tau)|_{c} \equiv - \langle n^{A}_{i}(\tau) n^{A}_{j}(0) \rangle
\end{equation}
such that, Operator $n^{A}_{i}= c^{\dag A\uparrow}_{i}c^{A\uparrow}_{i}+c^{\dag A \downarrow}_{i}c^{A\downarrow}_{i}- n $ where $n$ is the average number of occupancy. Similarly, we define spin susceptibility as,
\begin{equation}\label{eq:spin_sus_tau}
    \chi^{AB}_{ij}(\tau)|_{s} \equiv - \langle \mathcal{S}^{A,+}_{i}(\tau) \mathcal{S}^{B,-}_{j}(0) \rangle
\end{equation}
where, $\mathcal{S}^{A,+}_{i} = c^{\dag A\uparrow}_{i} c^{A\downarrow}_{i} $,  $\mathcal{S}^{A,-}_{i} = c^{\dag A \downarrow}_{i} c^{A,\uparrow}_{i}$.
In a non-interacting system with a paramagnetic state, we can apply Wick's Theorem and express the charge Eq. (\ref{eq:charge_sus_tau}) and spin susceptibility Eq. (\ref{eq:spin_sus_tau}) in terms of the product of two single-particle Green's function such that (see \ref{App:Susceptibility}) 
\begin{equation}
         \chi^{AB}_{ij}(\tau)|_{\mathbf{s}} = \frac{1}{2} \chi^{AB}_{ij}(\tau)|_{\mathbf{c}} =  \chi^{AB}_{0\,\,ij}(\tau)  \equiv  -\mathcal{G}_{ij}(\tau)\mathcal{G}_{ji}(-\tau)
\end{equation}
where $\chi^{AB}_{0\,\,ij}$ is defined as Lindhard susceptibility. Fourier transform of Lindhard susceptibility is given by,
\begin{equation}\label{eq:Lindhard_susceptibilty}
\begin{split}
 \chi_{0}^{AB}(\mathbf{q},\omega)  =  -\frac{1}{\beta} \sum_{\mathbf{k}} \sum_{\omega_n} \mathcal{G}^{A-B}(\mathbf{k}, i\omega_n) \mathcal{G}^{B-A}(\mathbf{k+q}, i\omega_n + \omega )
\end{split}
\end{equation} 
where $i\omega_{n}$ is Matsubara frequency, $\omega$ is the bosonic frequency, $\beta$ is the inverse temperature, and $ \mathcal{G}^{A-B}(\mathbf{k}, i\omega_n)$ is the single-particle Green’s function for orbital indices A and B. This is, in principle, a response of material to an external field, the peak of which also gives the optimal direction of momentum-dependent fluctuation/instabilities or potential ordering.

\begin{figure*}[!htbp]
    \centering
    \includegraphics[width=1\linewidth]{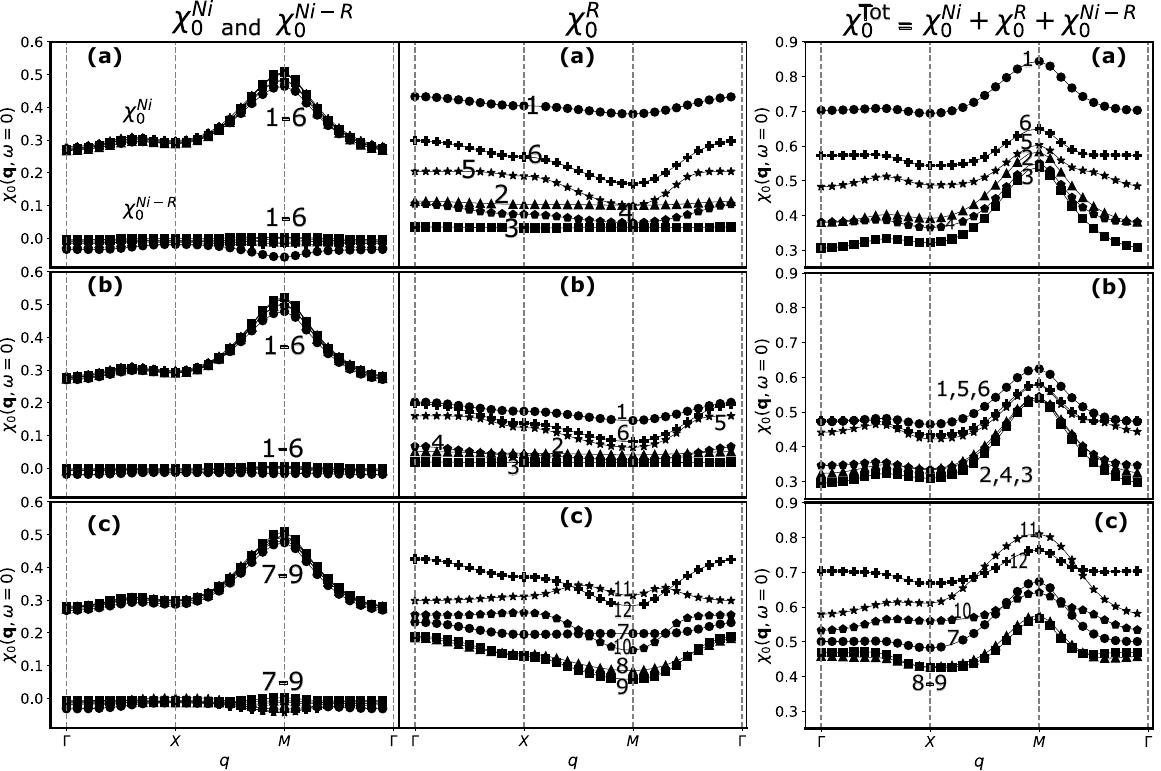}
    \caption{Lindhard susceptibility  components of $RNiO_{2}$ with condition \textbf{(a)}  $\Tilde{\varepsilon}_{3d_{z^2}}=\Tilde{\varepsilon}_{3d_{x^2-y^2}}$,  \textbf{(b)} $\Tilde{\varepsilon}_{3d_{z^2}}<\Tilde{\varepsilon}_{3d_{x^2-y^2}}$,  \textbf{(c)} $\Tilde{\varepsilon}_{3d_{z^2}}>\Tilde{\varepsilon}_{3d_{x^2-y^2}}$ (see Fig. \ref{fig:RNiOn_results}) following parameters with labels mentioned in Table \ref{tab:trend} (see also Fig. \ref{fig:Trend_OT}) from our proposed effective model Eq. (\ref{eq:two_band_dispersion_RNiO_2}). Left panels $\chi_{0}^{Ni}$ and $\chi_{0}^{Ni-R}$, Center panels $\chi_{0}^{R}$ and Right panels $\chi_{0}^{Tot} = \chi_{0}^{Ni} + \chi_{0}^{R} + \chi_{0}^{Ni-R}$.}
    \label{fig:Chi_Evol}
\end{figure*}

Similar to previous section \ref{RNiOn} we computed Lindhard susceptibility $\chi_{0}(\mathbf{q},\omega = 0)$ for three different condition Fig. \ref{fig:Chi_Evol}\textbf{(a)} $\Tilde{\varepsilon}_{3d_{z^2}}=\Tilde{\varepsilon}_{3d_{x^2-y^2}}$,  Fig. \ref{fig:Chi_Evol}\textbf{(b)} $\Tilde{\varepsilon}_{3d_{z^2}}<\Tilde{\varepsilon}_{3d_{x^2-y^2}}$ and Fig. \ref{fig:Chi_Evol}\textbf{(c)} $\Tilde{\varepsilon}_{3d_{z^2}}>\Tilde{\varepsilon}_{3d_{x^2-y^2}}$. Again these three condition mimic the effects of dopants(like Oxygen) and applied pressure as depicted in Figure. \ref{fig:RNiOn_results}. Using the proposed effective model given by the dispersion equation (\ref{eq:two_band_dispersion_RNiO_2}) for each of these cases, we tune the parameters depicted by labels $1-12$ shown  Table \ref{tab:trend}(or Fig.\ref{fig:Trend_OT}). These parameters are intentionally selected to take into account all possible variations of the tunable parameter to check its effect on  $\chi_{0}(\mathbf{q},\omega = 0)$ with the emergence of secondary Fermi-pocket. To understand the trend, consider the plotted $\chi_{0}(\mathbf{q},\omega = 0)$ in Fig.\ref{fig:Chi_Evol} for different parameters with labels as mentioned in Table \ref{tab:trend}. When $\Tilde{\varepsilon}_{5d_{z^2}}$ varies with a constant $t_{RR}$, $\chi_{0}^{R}$  sequentially decreases with an increasing $\Tilde{\varepsilon}_{5d_{z^2}}$.  This trend of $\chi_{0}^{R}$ is similar for all three conditions $\Tilde{\varepsilon}_{3d_{z^2}}=\Tilde{\varepsilon}_{3d_{x^2-y^2}}$, $\Tilde{\varepsilon}_{3d_{z^2}}<\Tilde{\varepsilon}_{3d_{x^2-y^2}}$, and $\Tilde{\varepsilon}_{3d_{z^2}}>\Tilde{\varepsilon}_{3d_{x^2-y^2}}$. When $t_{RR}$ varies with a constant $\Tilde{\varepsilon}_{5d_{z^2}}$, $\chi_{0}^{R}$  sequentially increases with an increasing $t_{RR}$ This trend of $\chi_{0}^{R}$ is also similar for all three cases mentioned earlier. Finally, when both $\Tilde{\varepsilon}_{5d_{z^2}}$ vary with $t_{RR}$, $\chi_{0}^{R}$  sequentially increases with a decreasing  $\Tilde{\varepsilon}_{5d_{z^2}}$  and $t_{RR}$. This $\chi_{0}^{R}$ trend is consistent over all three cases. It is evident from plotted Fig. \ref{fig:Chi_Evol} that this trend of $\chi_{0}^{R}$ becomes more dormant when $\Tilde{\varepsilon}_{3d_{z^2}}<\Tilde{\varepsilon}_{3d_{x^2-y^2}}$. Further, $\chi_{0}^{Ni-R}$ is consistently near $0$ throughout all the cases. $\chi_{0}^{Ni}$ follows the same trend as $\chi_{0}^{R}$ but is less robust compared to  $\chi_{0}^{R}$ and shows a consistent peak at $\mathbf{q}_{0}=(\pi,\pi)$, which has been explicitly observed through reported \textit{ab-initio} calculations \cite{Chen2022Nov,Zhou2020Aug,Liu2024Jun} and a small crest at $\mathbf{q}_{0}=(\pi,0)$ throughout all the cases. For the variation of $\Tilde{\varepsilon}_{5d_{z^2}} - t_{RR}$, $\chi_{0}^{R}$ shows a consistent peak at $\mathbf{q}_{0}=(0,0)$.

\begin{table*}[!htbp]
\centering
\renewcommand{\arraystretch}{1.5} 
\setlength{\tabcolsep}{10pt} 
\begin{tabular}{|l|c|c|c|}
\hline
\diagbox{$t_{RR}$}{$\Tilde{\varepsilon}_{5d_{z^2}}$} & 0.10 & 0.325 & 0.55 \\
\hline
1.00 &  \textbf{11} & \textbf{10} & \textbf{9} \\ \hline
0.775 & \textbf{12} & \textbf{8} &   \textbf{5} \\ \hline
0.55 &\textbf{7 } &  \textbf{6} &   \textbf{4} \\ \hline
0.325 &  \textbf{1}&   \textbf{2} &   \textbf{3} \\ \hline
\end{tabular}
\caption{Table of parameters followed for calculating $\chi_{0}(\mathbf{q},0)$ in Fig. \ref{fig:Chi_Evol} for $RNiO_{2}$ from proposed effective model given by dispersion Eq. (\ref{eq:two_band_dispersion_RNiO_2}). All values are in eV}
\label{tab:trend}
\end{table*}

Instabilities in the spin or charge channels can be understood by interacting susceptibilities, $\chi(\mathbf{q}, \omega)|_s = \chi_{0}(\mathbf{q},\omega)/(1 + V(\mathbf{q})_{s} \chi_{0}(\mathbf{q},\omega))$ for spin and $\chi(\mathbf{q}, \omega)|_c = \chi_{0}(\mathbf{q},\omega)/(1 - V(\mathbf{q})_{c} \chi_{0}(\mathbf{q},\omega))$ for charge in RPA formalism. Where $\chi_0(\mathbf{q}, \omega)$ is the Lindhard susceptibility and $V(\mathbf{q})_{s}$ and $V(\mathbf{q})_{c}$ are the effective interaction strength. In the spin channel, a divergence in $\chi(\mathbf{q},\omega)$ suggests instabilities towards magnetic ordering, such as the formation of spin density wave (SDWs)  and $V(\mathbf{q})_{s}$ corresponds to effective spin-spin interaction such as Heisenberg or Hund type exchange interaction and the peak in  $\chi(\mathbf{q},\omega)$ corresponds to the nesting wave vector where SDW formation is energetically favourable. In the charge channel, divergence indicates a tendency towards charge ordering, such as the formation of charge-density waves (CDWs). Peaks in $\chi_0$ correspond to wave vectors favouring SDW or CDW formation. In the charge channel, the $V(\mathbf{q})_{c}$ corresponds to effective charge-charge interaction such as Coulomb exchange interaction and the peak in  $\chi(\mathbf{q},\omega)$ corresponds to the nesting wave vector where electronic density exhibits a modulation in the direction of specific wave-vector. Thus, the RPA takes into account how interactions affect the Lindhard susceptibility, resulting in an increased response at specific wave vectors $(\mathbf{q}_{0})$ and frequencies $(\omega)$ where collective excitations or instabilities may occur. When the Lindhard susceptibility is sufficiently large ($\chi_0(\mathbf{q},\omega) \sim V(\mathbf{q})_{s}^{-1} \text{\,\,or\,\,}  V(\mathbf{q})_{c}^{-1}$), the system becomes prone to long-range ordering. In our model, considering the spin channel, an increase in interlayer hybridization is observed when $\Tilde{\varepsilon}_{3d_{z^2}}<\Tilde{\varepsilon}_{3d_{x^2-y^2}}$. In this scenario, we observe a dominant contribution of $\chi_{0}^{Ni}$ near $\mathbf{q}_{0} = (\pi, \pi)$ (the $M$-point) compared to a relatively inactive $\chi_{0}^{R}$ in $\chi_{0}^{Tot}$. This suggests the emergence of an antiferromagnetic-ordered state, similar to Cuprates \cite{Vladimirov2011Jan, Sidis2004May, Sidis2007, Eschrig2006Jan, Si1993Apr, Fong2000Jun, Dai2001Jan, Yamase2006Jun}.In Cuprates, doping rapidly disrupts this long-range order, indicating the emergence of short-range magnetic correlations due to magnetic instabilities. This shift in magnetic behaviour underscores the potential role of spin fluctuations in the electronic pairing mechanisms \cite{Singh2021Jan}.  A similar emergence has been reported in determinant quantum Monte Carlo (DQMC) studies in Layered Nickelates. Furthermore, in layered Nickelates, the reduction in the dimensionality of the magnetic fluctuations (from 3D to 2D) coincides with the emergence of superconductivity, which aligns with our assumptions and results for the proposed effective model. However, on the other hand, $\chi_{0}^{R}$ shows an emergence of a peak at $\mathbf{q}_{0} = (0, 0)$ (the $\Gamma$-point) when the energy of the $\Tilde{\varepsilon}_{3d_{z^2}}$ orbital exceeds or equals that of the $\Tilde{\varepsilon}_{3d_{x^2-y^2}}$ orbital. This suggests the emergence of ferromagnetic ordering, which is highly unlikely to be the origin of the emergence of any superconducting order. It is noted that there are several theoretical and experimental attempts to investigate the pairing mechanism \cite{Sun2023Sep, Zhang2023Oct, Ryee2024Aug} in $R_{3}Ni_{2}O_{7}$ and elevating the $T_c$ in $Pr_{1-x}Sr_{x}NiO_{2}$ \cite{Wang2022Jul, DiCataldo2024May}. 

The peak in $\chi_{0}^{Tot}$ is generally attributed to ordering tendency or instabilities-whether spin or charge. These peaks are typically associated with Fermi surface nesting or other mechanisms that enhance the electronic response at specific wave vectors. In our model for layered Nickelates, increasing pressure enhances the ordering tendency and favors the formation of a single-pocket Fermi surface, consistent with recent experimental results \cite{DiCataldo2024May}. Similarly, optimal doping—whether electron or hole doping—also strengthens the overall tendency for ordering and the appearance of a single-pocket Fermi surface, as confirmed by recent ARPES measurements \cite{Ding2024Mar, Sun2024Mar, Si2024Aug}. Our case of $\Tilde{\varepsilon}_{3d_{z^2}}=\Tilde{\varepsilon}_{3d_{x^2-y^2}}$ corresponds to electron doping which can induce a phase transition to a structurally ordered state. For example, excess oxygen doping in layered Nickelates may result in a transition to a complete perovskite phase, which would explain the observation of a two-pocket Fermi surface resembling that of perovskite phases \cite{Lee2011Oct, Zhong2024Jul}. The complementary explanations could be that the incomplete reduction can lead to residual ordering as observed in $Ni-L_{3}$ edge resonant X-ray scattering experiment \cite{Tam2022Oct}, whereas complete reduction to the layered phase shows no evidence of any ordered phase\cite{Parzyck2024Apr}. 

Thus, from our proposed model, we computed the Lindhard susceptibility $\chi_0(\mathbf{q},0)$ for various energy configurations to simulate the effects of doping and applied external pressure qualitatively. Doping and applied external pressure can be attributed to the competition between different ordering wave vectors originating from Rare-earth and Nickel. Similar to $CuO_{2}$ layer in Cuprates superconductors, $NiO_{2}$ also has tendency of $\mathbf{q}_{0} = (\pi, \pi)$ ordering. However, depending on external factors such as doping or pressure, Rare-earth might compensate for this order. We also discussed the versatility of our model to aid the understanding of electronic charge or spin instabilities in layered Nickelates. 

\section{Conclusion}

This study provides a comprehensive modelling framework for understanding the role of effective orbitals and the emergence of an additional Fermi pocket in layered Nickelates. The analysis shows that the Nickel $3d_{x^2-y^2}$ orbital mainly crosses the Fermi level, while the Nickel $3d_{z^2}$ orbital acts as an anchor to lowering the energy of the Rare-earth $5d_{z^2}$ orbital, leading to the emergence of a second Fermi surface pocket. This anchoring effect of $3d_{z^2}$ allows an effective two-orbital description involving both Nickel and Rare-earth, explaining how the Fermi surface emergence evolves with doping and pressure qualitatively-similar to features seen in Cuprate superconductors.

Based on our observations in Lindhard susceptibility calculations, we cannot confirm any instabilities at $\mathbf{q}_{0} = (\frac{1}{3},0)$ in reciprocal lattice units (r.l.u), as suggested by resonant X-ray scattering experiment \cite{Tam2022Oct}. However, our findings on susceptibility indicate potential instabilities at $\mathbf{q}_{0} = (\pi, \pi)$ (or $\mathbf{q}_{0} = (\frac{1}{2}, \frac{1}{2})$ in r.l.u), similar to those observed in Cuprates. The peak in Lindhard susceptibility could be attributed to the strong influence of the underlying square lattice symmetry, similar to Cuprates and is tunable with pressure and doping. Specific interaction terms must be considered to understand the microscopic pairing mechanism for superconductivity, whether it involves CDW, charge order, or magnetic order. This sheds light on the most relevant interaction and provides insights into the microscopic pairing mechanism for superconductivity. The observed Lindhard susceptibility peak at $\mathbf{q}_{0} = (\pi, \pi)$ will be crucial for the emergence of any ordering instability, including superconductivity, resulting from microscopic interaction terms. The proposed modelling framework provides a comprehensive approach to studying the effects of doping and pressure in the Nickelate family. It offers insights into the factors influencing susceptibilities, such as Fermi-surface nesting, which significantly impact the fluctuation spectrum.

The proposed two-orbital effective model for layered Nickelates could be used in further work incorporating electronic interactions more accurately, which will provide deeper insights into phenomena such as charge and spin density waves, ordering, and other instabilities contributing to superconductivity. Further investigation of varying doping levels and external pressures, informed by empirical data and \textit{ab initio} calculations, is also recommended. Experimental studies of $(\pi,\pi)$ instabilities in the spin sector, such as spin-density waves and N{\'e}el order, and in the charge sector, like charge-density waves, could offer crucial evidence. If the origin and presence or absence of correlations are confirmed experimentally, they would significantly enhance our understanding of the emergence of superconductivity in Nickelates and their connection to Cuprates.

\section*{ACKNOWLEDGMENTS}
This work has been supported by Centre national de la recherche scientifique (CNRS) and Agence National de Recherche (ANR) grant SuperNickel (ANR-21-CE30-0041-02)

\newpage

\section*{Supplementary Material: Emergence and tunability of Fermi-pocket and electronic instabilities in layered Nickelates}
\subsection*{Effect of Three dimensionality}\label{App:3D}
Three-dimensionality plays a crucial role in understanding perovskite Nickelates' electronic and structural properties. These materials exhibit rich phase diagrams and intriguing phenomena such as metal-insulator transitions and high-temperature superconductivity. The three-dimensional nature of the crystal structure and electronic interactions introduces complexity, especially when considering the interactions between atoms along different crystallographic axes. However, in many cases, this complexity can be effectively captured by approximations that simplify the problem. One such approach is to fix the $ k_z $ momentum component, assuming only the on-site energy is renormalized. This simplification reduces the computational burden while preserving essential features of the system’s three-dimensional behavior. It allows for a more tractable analysis of their electronic properties without sacrificing key physical insights.

Recent progress in superconductivity studies of layered Nickelates such as $ RNiO_2 $ has highlighted the importance of these three-dimensional considerations for consistently synthesizing the parent perovskite compound $ RNiO_3 $ and its subsequent reduction to $ RNiO_2 $ \cite{Li2019Aug}. The transformation from perovskite $ RNiO_3 $ to the superconducting $ RNiO_2 $ phase, such as $ LaNiO_2 $, $ NdNiO_2 $, and $ PrNiO_2 $, involves the selective removal of apical Oxygen, which alters the electronic structure and dimensionality. These layered Nickelates exhibit superconductivity when hole-doped, typically by substituting rare-earth elements with $ Sr $ or $ Ca $ \cite{Zeng2020Oct, Zeng2022Feb, Osada2020Dec, Osada2021Nov, Li2019Aug, Li2020Jul}.

In perovskite Nickelates, the Fermi surface is predominantly shaped by the $ 3d_{x^2-y^2} $ and $ 3d_{z^2} $ orbitals of Nickel, with electron hopping occurring through two primary mechanisms: an in-plane channel involving the $ 3d_{x^2-y^2} $ orbital hybridizing with Oxygen $ 2p_{x} $ and $ 2p_{y} $, and an out-of-plane channel where the $ 3d_{z^2} $ orbital hybridizes with apical Oxygen $ 2p_{z} $ \cite{Dhaka2015Jul}. Additional contributions from rare-earth $ 5d_{z^2} $ orbitals further complicate the three-dimensional electronic structure, which must be accounted for when constructing a generalized Hamiltonian.

This generalized Hamiltonian incorporates key orbitals—rare-earth $ 5d_{z^2} $, Nickel $ 3d_{x^2-y^2} $ and $ 3d_{z^2} $, and Oxygen $ 2p_{x/y/z} $—and models electron hopping through neighboring Oxygen $ p $-states as a virtual process. The block-diagonal form of the Hamiltonian reflects the non-degenerate on-site energies of the transition metal and rare-earth orbitals, while the degenerate, lower-energy Oxygen $ p $-orbitals preserve key symmetries. By fixing the $ k_z $ component and focusing on on-site energy renormalization, the dimensional reduction captures essential three-dimensional effects, enabling a unified description of both perovskite $ RNiO_3 $ and superconducting $ RNiO_2 $, without losing the characteristic emergence of the two-pocket Fermi surface.

Thus, the blocks in Eq. (\ref{eq:Generalized_Dispersion_k_matrix}) are given by,
    \begin{subequations}\label{Eq:GeneralizedDispersion$RNiO_{3}$}
    \begin{align}
        \Tilde{\varepsilon}^{O}_{\mathbf{k}}  = \Tilde{\varepsilon}_{2p} 
        +\left[\begin{array}{cc}
            0 &  t_{pp} s_{k_x} s^{*}_{k_y}\\
            t_{pp} s^{*}_{k_x} s_{k_y}  & 0 \end{array}\right] 
        + \left[\begin{array}{cc}
            0 &  t^{\perp}_{pp} s_{k_z}\\
            t_{pp}^{\perp} s^{*}_{k_z}  & 0 \end{array}\right]\label{Eq:GeneralizedDispersion$RNiO_{3}$_O}\\
       \Tilde{\varepsilon}^{NiO}_{\mathbf{k}}  = \left[\begin{array}{cc}
            -t^{\prime}_{pd}s_{k_x} & -t^{\prime}_{pd}s_{k_y} \\
            t_{pd} s_{k_x} & -t_{pd} s_{k_y}  \end{array}\right]  
        +\left[\begin{array}{cc}
            0 & t^{\prime\perp}_{pd}s_{k_z}\\
            0 & 0 \end{array}\right] 
            \label{Eq:GeneralizedDispersion$RNiO_{3}$_Ni_O}\\
      \Tilde{\varepsilon}^{Ni}_{\mathbf{k}}  =  \left[\begin{array}{cc}
            \Tilde{\varepsilon}_{3d_{z^2}} & 0 \\
            0 & \Tilde{\varepsilon}_{3d_{x^2-y^2}} \\ \end{array}\right] 
        + \left[\begin{array}{cc}
            0 &  t_{dd}\\
            t_{dd} & 0 \\ \end{array}\right]
        + \left[\begin{array}{cc}
            0 &  t^{\perp}_{dd} s_{k_z}\\
            t^{\perp}_{dd} s^{*}_{k_z} & 0 \\ \end{array}\right]\label{Eq:GeneralizedDispersion$RNiO_{3}$_Ni}\\
        \Tilde{\varepsilon}^{R}_{\mathbf{k}}  = \Tilde{\varepsilon}_{5d_{z^2}} + t_{RR}(\cos(k_x)+\cos(k_y)) + t^{\perp}_{RR} \cos(k_z) \\
        \Tilde{\varepsilon}^{RNi}_{\mathbf{k}}  = \left[\begin{array}{cc}
            t^{\prime}_{dd}s_{k_x}s_{k_y}  & t^{\prime\prime}_{dd}s_{k_x}s_{k_y}\end{array}\right] + \left[\begin{array}{cc}
            t^{\prime\perp}_{dd}s_{k_z}  & 0 \end{array}\right]  \label{Eq:GeneralizedDispersion$RNiO_{3}$_R_Ni}
    \end{align}
\end{subequations}
where 
\begin{equation}
 s_{k_{\nu}} =(1-e^{i k_{\nu}})
\end{equation}
and its complex conjugate $s_{k_{\nu}}^* =(1-e^{-i k_{\nu}})$.

\begin{figure}[!htbp]
    \centering
    \includegraphics[width=0.7\linewidth]{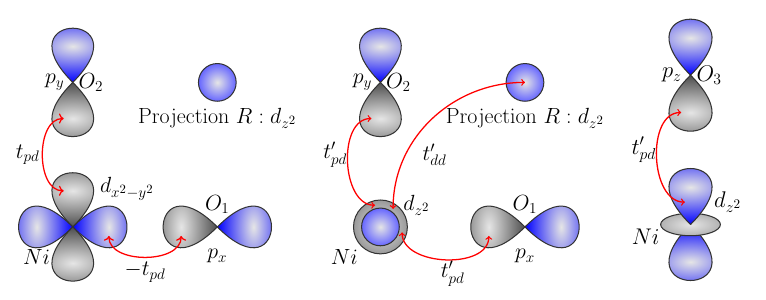}
    \caption{Orbital overlaps and the hopping parameter considering the orbital symmetry of $RNiO_{n}$ and the in-plane Oxygen atoms $O_{1}$ and $O_{2}$ as well as the out-of-plane Oxygen atom $O_{3}$. }
    \label{fig:Orbital_Overlaps}
\end{figure}
\begin{figure}[!htbp]
    \centering
    \includegraphics[width=0.5\linewidth]{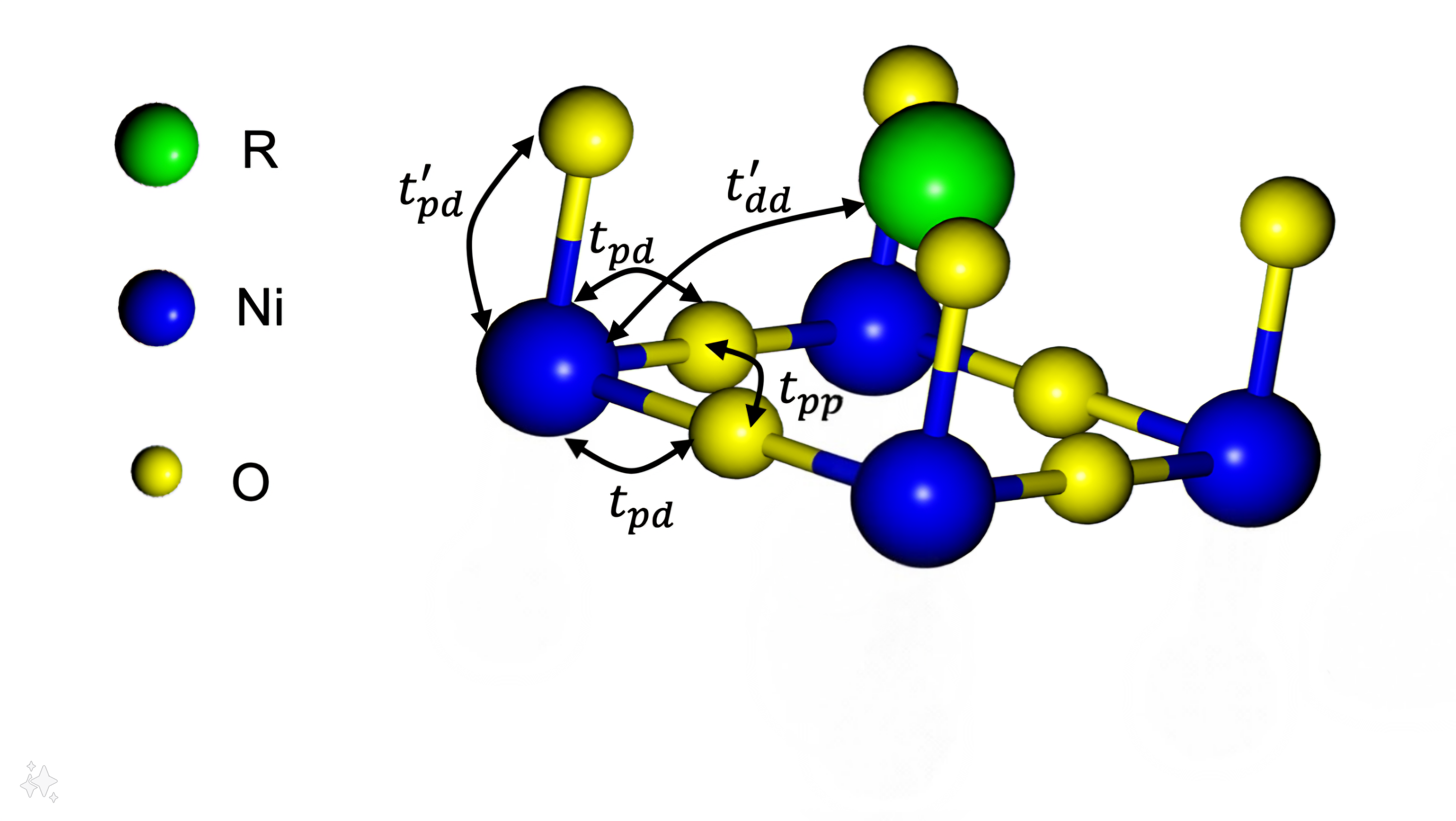}
    \caption{Schematic of in-plane and out-plane hopping parameter in $RNiO_{3}$.}
    \label{fig:$RNiO_{3}$}
\end{figure} 

As shown in Figs. \ref{fig:$RNiO_{3}$} and \ref{fig:Orbital_Overlaps}, the parameter labeled as $t_{pd}$ represents the hopping parameter between $3d_{x^2-y^2}-2p_{x/y}$ in the $x-y$ plane. $t^{\prime}_{pd}$ denotes the hopping parameter between the $3d_{z^2}-2p_{z}$ out of the $x-y$ plane. The parameters $t^{\prime}_{dd}$, $t^{\prime\prime}_{dd}$ correspond to the hopping parameters between $5d_{z^2}-3d_{z^2}$ and $5d_{z^2}-3d_{x^2-y^2}$ respectively. The $t_{dd}$ represents the hybridization between $3d_{z^2}-3d_{x^2-y^2}$. Finally, $t_{pp}$ denotes the hopping parameter between $2p_{x}-2p_{y}$ in the $x-y$ plane, and $t_{RR}$ is the hopping between the Rare-earth's $5d_{z^2}$. 
Further, the parameters $t^{\perp}_{pp},t^{'\perp}_{pd}$ and $t_{dd}^{'\perp}$ corresponds to out of plane (perpendicular) component of $2p_{x/y}-2p_{z},3d_{z^2}-3p_{z}$ and $5d_{z^2}-3d_{z^2}$ hopping respectively and $t^{\perp}_{dd},t^{\perp}_{RR}$ is out of plane (perpendicular) component of hybridization between $Ni-Ni$ and $R-R$ orbitals. At a fixed value of $k_{z}$, these $t^{\perp}$'s renormalize the on-site contribution of the corresponding orbital. In our calculation, we fix this $k_{z} = 0$ and assume the same fixed parameter as of layered Nickelates except for the degeneracy of $e_g$ Indeed, this is over-simplification, but the aim here is explicitly limited to proposing the model for perovskite structure in the spirit, of same assumption as for layered structure.

\begin{figure*}[!htbp]
    \centering
    \includegraphics[width=0.85\linewidth]{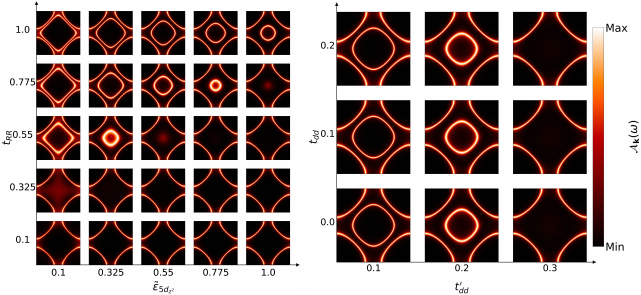}
    \caption{Fermi surface evolution in $RNiO_{3}$. Left Panel: Varying $t_{RR}$ and $\Tilde{\varepsilon}_{5d_{z^2}}$ for a  fixed $t^{\prime}_{dd} = 0.2\,eV$ and $\Tilde{\varepsilon}_{3d_{z^2}}= \Tilde{\varepsilon}_{3d_{x^2-y^2}} = -1.0\,eV$. Right panel: Varying $t^{\prime}_{dd}$ and $t_{dd}$ for a fixed $t_{RR} = 0.775 \,eV$ and $\Tilde{\varepsilon}_{5d_{z^2}} = 0.55 \,eV$, with $\Tilde{\varepsilon}_{3d_{z^2}} = \Tilde{\varepsilon}_{3d_{x^2-y^2}} = -1.0\,eV$. Each Fermi-surface extend from $-\pi$ to $\pi$ in $(k_x,k_y)$ plane.}
    \label{fig:FS_Evol_1.0}
\end{figure*}

Further, in perovskite $RNiO_3$ compounds, the octahedral crystal field induces a level splitting such that $t_{2g} < e_g$, leaving the $3d_{z^2}$ and $3d_{x^2-y^2}$ orbitals degenerate. Our model accounts for the symmetry constraints that prevent orbital overlap between the $5d_{z^2}$ and $2p_{x/y/z}$ orbitals, as well as between the $2p_z$ and in-plane $2p_{x/y}$ and $3d_{x^2-y^2}$ orbitals. To examine the effect of apical Oxygen on orbital energies within our multi-orbital model of $RNiO_3$, described by the dispersion in Eq. (\ref{Eq:GeneralizedDispersion$RNiO_{3}$}), we analyze the interplay among the $5d_{z^2}$, $3d_{x^2-y^2}$, $3d_{z^2}$, and $2p_{x}$, $2p_{y}$, $2p_{z}$ orbitals. By adjusting the relative energy scales and hopping parameters, we aim to tune the model to capture the critical emergence of an additional Fermi pocket using a spectral function Eq. (\ref{eq:GF}) derived from Green's function formalism. This parametrization tuning results in a Fermi surface that spans the Brillouin zone from antinodal to nodal regions. Notably, there is a varying energy gap along the antinodal direction compared to the nodal direction, reflecting a substantial pocket centered at $\Gamma (0,0)$, extending from $-\pi$ to $\pi$ in the high-symmetry plane $\Gamma(0,0)-X(\pi,0)-M(\pi,\pi)-\Gamma(0,0)$, as shown in Fig. \ref{fig:FS_Evol_1.0}.

Thus, the Fermi surface evolution depicted in Fig. \ref{fig:FS_Evol_1.0} explores the versatility of our 6-band model's behavior, with dispersion given by Eq. (\ref{Eq:GeneralizedDispersion$RNiO_{3}$}). By varying different parameters, it elucidates the interdependence of the $3d_{z^2}$ and Rare-earth $5d_{z^2}$ orbitals and the resulting formation of one or two Fermi surface pockets. As we have considered degenerate $e_g$ near the Fermi level, one may expect the direct contribution of these levels on the Fermi surface. However, as aforementioned due to symmetry reasons $3d_{z^2}$ rather overlap significantly with $5d_{z^2}$ causing extended orbitals. Hopping through these extended orbitals results in the second pocket in the Fermi surface. This pocket is robust over variation of hopping amplitude between $5d_{z^2}-3d_{z^2}$ but dormant with variation in hopping amplitude between $3d_{z^2}-3d_{x^2-y^2}$.  Thus, we can safely include the $3d_{z^2}$ as the anchoring orbital for proposing the two orbital effective model, discussed in detail in section \ref{RNiOn} in main text.
\newpage
\appendix
\newpage
\section{Downfolding}\label{App:Downfolding}
This method of reduction is permitted because of low-lying Oxygen. In principle, we can safely downfold (integrate out) Oxygen's contribution to the dispersion, yielding an effective Hamiltonian. This new downfolded Hamiltonian is restricted to the low-energy description. For full Hamiltonian of form of Eq.(\ref{eq:Generalized_Hamiltonian_k}) the Green's function matrix have form of, 
\begin{equation}
\mathcal{G}_{\mathbf{k}} (\omega) = \left[\begin{array}{c|c|c}
\mathcal{G}^{R}_{\mathbf{k}} (\omega) & \mathcal{G}_{\mathbf{k}}^{RNi} (\omega)& \mathcal{G}_{\mathbf{k}}^{RO} (\omega)\\
&&\\
\hline
&&\\
\mathcal{G}_{\mathbf{k}}^{NiR} (\omega) & \mathcal{G}^{Ni}_{\mathbf{k}}(\omega) & \mathcal{G}_{\mathbf{k}}^{NiO} (\omega)\\ 
&&\\
\hline
&&\\
\mathcal{G}_{\mathbf{k}}^{OR} (\omega)& \mathcal{G}_{\mathbf{k}}^{ONi} (\omega) & \mathcal{G}^{O}_{\mathbf{k}} (\omega)
    \end{array}\right]
\end{equation}
with dimension $N\times N$. Where, each element $\mathcal{G}_{\mathbf{k}}(\omega) = (\omega-\Tilde{\varepsilon}_{\mathbf{k}})^{-1}$. Upon downfolding Oxygen contribution, we can have the effective Green's function matrix of size $(N-N_{O})\times(N-N_{O})$ such that,

\begin{equation}\label{eq:eff_GF_1}
\mathcal{G}_{eff}(\mathbf{k},\omega) = \left[\begin{array}{c|c}
\mathcal{G}^{R}_{\mathbf{k}} (\omega) & \mathcal{G}_{\mathbf{k}}^{RNi} (\omega) \\
&\\
\hline
&\\
\mathcal{G}_{\mathbf{k}}^{NiR} (\omega) & \mathcal{G}^{Ni}_{\mathbf{k}} (\omega)\end{array}\right]
\end{equation}
where we introduce self energy contribution $\Sigma(\mathbf{k},\omega)$ such that,

\begin{equation}
\begin{split}
    \bigg[ \mathcal{G}_{eff}(\mathbf{k},\omega) \bigg]^{-1} & =  \omega \mathbb{1} -   \left[\begin{array}{c|c}
\Tilde{\varepsilon}^{R}_{\mathbf{k}} & \Tilde{\varepsilon}^{RNi}_{\mathbf{k}} \\
&\\
\hline
&\\
\Tilde{\varepsilon}^{NiR}_{\mathbf{k}} & \Tilde{\varepsilon}^{Ni}_{\mathbf{k}}\end{array}\right] +  \left[\begin{array}{c|c}
\Sigma^{R}_{\mathbf{k}}(\omega) & \Sigma^{RNi}_{\mathbf{k}}(\omega) \\
&\\
\hline
&\\
\Sigma^{NiR}_{\mathbf{k}}(\omega) & \Sigma^{Ni}_{\mathbf{k}}(\omega) \end{array}\right] \\ & = (\omega \mathbb{1} - \mathcal{E}_{eff}(\mathbf{k},\omega))  \approx (\omega \mathbb{1} - \mathcal{E}_{eff}(\mathbf{k})) 
\end{split}
\end{equation}

such that,
\begin{subequations}
    \begin{align}
        \Sigma^{R}_{\mathbf{k}}(\omega) & = \Tilde{\varepsilon}_{\mathbf{k}}^{RO}  [\omega - \Tilde{\varepsilon}^{O}_{\mathbf{k}}]^{-1} \Tilde{\varepsilon}_{\mathbf{k}}^{OR}\\
        \Sigma^{Ni}_{\mathbf{k}}(\omega) & =  \Tilde{\varepsilon}_{\mathbf{k}}^{NiO}  [\omega - \Tilde{\varepsilon}^{O}_{\mathbf{k}}]^{-1} \Tilde{\varepsilon}_{\mathbf{k}}^{ONi} \\
       \Sigma^{RNi}_{\mathbf{k}}(\omega) & = \Tilde{\varepsilon}_{\mathbf{k}}^{RO}  [\omega - \Tilde{\varepsilon}^{O}_{\mathbf{k}}]^{-1} \Tilde{\varepsilon}_{\mathbf{k}}^{ONi}
    \end{align}
\end{subequations}
On the general ground, this method of downfolding is exact since we do not consider interaction in our model for the Oxygen orbitals. The self energies are $\omega-$ dependent, however we assume that the energies of $\Tilde{\varepsilon}^{O}_{\mathbf{k}}$ are far from the Fermi-level of $R$ and $Ni$ electrons. Therefore, we can neglect this $\omega$ dependence and approximate these self-energies by their static $(\omega=0)$ components. Thus, from effective Green's function Eq. (\ref{eq:eff_GF_1}) we can have effective dispersion of form, 

\begin{equation}\label{eq:two_band_dispersion}
\mathcal{E}_{eff}(\mathbf{k}) = \left[\begin{array}{c|c}
\mathcal{\Tilde{E}}_{1} (\mathbf{k}) & V_{12}(\mathbf{k}) \\
&\\
\hline
&\\
V_{12}^{*}(\mathbf{k}) & \mathcal{\Tilde{E}}_{2} (\mathbf{k})
\end{array}\right]
\end{equation}
such that,
\begin{subequations}\label{eq:explicit_self_energies}
    \begin{align}
        \mathcal{\Tilde{E}}_{1} (\mathbf{k})  & = \Tilde{\varepsilon}^{R}_{\mathbf{k}} + \Tilde{\varepsilon}_{\mathbf{k}}^{RO}  [\Tilde{\varepsilon}^{O}_{\mathbf{k}}]^{-1} \Tilde{\varepsilon}_{\mathbf{k}}^{OR} \\
        \mathcal{\Tilde{E}}_{2} (\mathbf{k})  & = \Tilde{\varepsilon}^{Ni}_{\mathbf{k}} + \Tilde{\varepsilon}_{\mathbf{k}}^{NiO}  [ \Tilde{\varepsilon}^{O}_{\mathbf{k}}]^{-1} \Tilde{\varepsilon}_{\mathbf{k}}^{ONi} \\
       V_{12}(\mathbf{k}) & = \Tilde{\varepsilon}_{\mathbf{k}}^{RNi}+  \Tilde{\varepsilon}_{\mathbf{k}}^{RO}  [ \Tilde{\varepsilon}^{O}_{\mathbf{k}}]^{-1} \Tilde{\varepsilon}_{\mathbf{k}}^{ONi}
    \end{align}
\end{subequations} 
This assumes that the valence fluctuation on Oxygen can be relatively small compared to the onsite Coulomb repulsion, which is the usual case in most layered Nickelates.

\section{Susceptibility}\label{App:Susceptibility}
\subsection{Non-interacting charge susceptibility}
We define the charge susceptibility as the response function that measures the correlation between charge densities at different lattice sites. It is given by
\begin{equation}
    \chi_{ij}(\tau)|_{\mathbf{c}} \equiv - \langle n_{i}(\tau) n_{j}(0) \rangle 
\end{equation}
where the charge density operator $n_i$ is defined as
\begin{equation}
    n_{i}= c^{\dag\uparrow}_{i}c^{\uparrow}_{i}+c^{\dag\downarrow}_{i}c^{\downarrow}_{i}-n
\end{equation}
Here, $c^{\dag\sigma}_i$ and $c^{\sigma}_i$ are the Fermionic creation and annihilation operators, respectively, for spin $\sigma$ at site $i$, and $n$ is the average number of particles per site.

Expanding the correlation function, we have
\begin{equation}
\begin{split}
    \chi_{ij}(\tau)|_{\mathbf{c}} & = - \langle c^{\dag\uparrow}_{i}c^{\uparrow}_{i}(\tau)c^{\dag\uparrow}_{j}c^{\uparrow}_{j}(0) \rangle - \langle c^{\dag\uparrow}_{i}c^{\uparrow}_{i}(\tau)c^{\dag\downarrow}_{j}c^{\downarrow}_{j}(0)\rangle \\ 
    & \quad - \langle c^{\dag\downarrow}_{i}c^{\downarrow}_{i}(\tau)c^{\dag\uparrow}_{j}c^{\uparrow}_{j}(0) \rangle  - \langle c^{\dag\downarrow}_{i}c^{\downarrow}_{i}(\tau)c^{\dag\downarrow}_{j}c^{\downarrow}_{j}(0) \rangle + n^2
\end{split}
\end{equation}

In the non-interacting case, we can apply Wick's theorem to express the four-point correlation functions as products of two-point correlation functions:

\begin{equation}
\begin{split}
    \chi_{ij}(\tau)|_{\mathbf{c}} 
    & = - \langle c^{\dag\uparrow}_{i}(\tau)c^{\uparrow}_{j}(0) \rangle \langle c^{\dag\uparrow}_{j}(0)c^{\uparrow}_{i}(\tau)\rangle 
    - \langle c^{\dag\uparrow}_{i}(\tau) c^{\downarrow}_{j}(0)\rangle \langle c^{\dag\downarrow}_{j}(0) c^{\uparrow}_{i}(\tau)\rangle \\
    & \quad - \langle c^{\dag\downarrow}_{i} (\tau) c^{\uparrow}_{j}(0) \rangle \langle c^{\dag\uparrow}_{j}(0)c^{\downarrow}_{i}(\tau)\rangle 
    - \langle c^{\dag\downarrow}_{i}(\tau)c^{\downarrow}_{j}(0) \rangle \langle c^{\dag\downarrow}_{j}(0)c^{\downarrow}_{i}(\tau)\rangle
\end{split}
\end{equation}

By introducing the Green's function $\mathcal{G}_{ij}^{\sigma\sigma'}(\tau)$, defined as
\begin{equation}
    \mathcal{G}_{ij}^{\sigma\sigma'}(\tau) = - \langle T_\tau c_i^\sigma(\tau) c_j^{\dag\sigma'}(0) \rangle
\end{equation}
where $T_\tau$ denotes time ordering in imaginary time, we can rewrite the charge susceptibility as
\begin{equation}
\begin{split}
    \chi_{ij}(\tau)|_{\mathbf{c}} 
    & = -\mathcal{G}^{\uparrow\uparrow}_{ij}(\tau)\mathcal{G}^{\uparrow\uparrow}_{ji}(-\tau) 
    - \mathcal{G}^{\uparrow\downarrow}_{ij}(\tau)\mathcal{G}^{\downarrow\uparrow}_{ji}(-\tau) \\
    & \quad - \mathcal{G}^{\downarrow\uparrow}_{ij}(\tau)\mathcal{G}^{\uparrow\downarrow}_{ji}(-\tau) 
    - \mathcal{G}^{\downarrow\downarrow}_{ij}(\tau)\mathcal{G}^{\downarrow\downarrow}_{ji}(-\tau)
\end{split}
\end{equation}

In the paramagnetic case, due to spin SU(2) symmetry, the Green's functions satisfy
\begin{equation}
    \mathcal{G}^{\sigma\sigma'}_{ij} = \delta_{\sigma\sigma'} \mathcal{G}_{ij}
\end{equation}
where $\mathcal{G}_{ij}$ is the spin-independent Green's function. Thus, the charge susceptibility simplifies to
\begin{equation}
   \chi_{ij}(\tau)|_{\mathbf{c}}  = -2\mathcal{G}_{ij}(\tau)\mathcal{G}_{ji}(-\tau)
\end{equation}

\subsection{Non-interacting spin susceptibility}
We define the transverse spin susceptibility as the correlation function for the transverse spin components, given by
\begin{equation}
    \chi^{+-}_{ij}(\tau)|_{\mathbf{s}} \equiv - \langle \mathcal{S}^{+}_{i}(\tau) \mathcal{S}^{-}_{j}(0) \rangle
\end{equation}
where the spin raising operator $\mathcal{S}^{+}_i$ and spin lowering operator $\mathcal{S}^{-}_i$ are defined as
\begin{equation}
    \mathcal{S}^{+}_{i} = c^{\dag\uparrow}_{i} c^{\downarrow}_{i}, \quad \mathcal{S}^{-}_{i} = c^{\dag\downarrow}_{i} c^{\uparrow}_{i}
\end{equation}
Therefore,
\begin{equation}
\begin{split}
  \chi^{+-}_{ij}(\tau)|_{\mathbf{s}}  & = - \langle \mathcal{S}^{+}_{i}(\tau) \mathcal{S}^{-}_{j}(0) \rangle \\ 
 & = - \langle c^{\dag\uparrow}_{i} c^{\downarrow}_{i}(\tau) c^{\dag\downarrow}_{j} c^{\uparrow}_{j}(0) \rangle
\end{split}
\end{equation}
Using Wick's theorem,
\begin{equation}
\begin{split}
     \chi^{+-}_{ij}(\tau)|_{\mathbf{s}} & = - \langle c^{\dag\uparrow}_{i}(\tau) c^{\uparrow}_{j}(0) \rangle \langle c^{\dag\downarrow}_{j}(0) c^{\downarrow}_{i}(\tau) \rangle \\
  & = -\mathcal{G}^{\uparrow\uparrow}_{ij}(\tau)\mathcal{G}^{\downarrow\downarrow}_{ji}(-\tau) \\ 
\end{split}
\end{equation}
Assuming the paramagnetic state (which implies $\mathcal{G}^{\sigma\sigma'}_{ij} = \delta_{\sigma\sigma'}\mathcal{G}_{ij}$ due to spin SU(2) symmetry), the spin susceptibility simplifies as follows:
\begin{equation}
     \chi^{+-}_{ij}(\tau)|_{\mathbf{s}}  = -\mathcal{G}_{ij}(\tau)\mathcal{G}_{ji}(-\tau)
\end{equation}
Where, we define the Lindhard susceptibility as,
\begin{equation}
    \chi_{0\,\,ij}(\tau) \equiv  -\mathcal{G}_{ij}(\tau)\mathcal{G}_{ji}(-\tau)
\end{equation}
Thus, in the non-interacting paramagnetic case, the transverse spin susceptibility relates to the charge susceptibility as
\begin{equation}
         \chi^{+-}_{ij}(\tau)|_{\mathbf{s}} = \frac{1}{2} \chi_{ij}(\tau)|_{\mathbf{c}} =  \chi_{0\,\,ij}(\tau)
\end{equation}

\subsection{Random Phase Approximation (RPA)} \label{App:RPA}

In the non-interacting case, we have derived the charge and spin susceptibilities, $\chi_{ij}(\tau)|_{\mathbf{c}}$ and $\chi^{+-}_{ij}(\tau)|_{\mathbf{s}}$, as functions of the non-interacting Green's function $\mathcal{G}_{ij}(\tau)$. However, when interactions are present, the susceptibilities are modified due to collective screening effects. The Random Phase Approximation (RPA) provides a way to account for these interaction effects perturbatively. Here, we will highlight some essential results from this well-known method for charge channel. 

In the presence of interactions, the full susceptibility can be expressed in terms of the non-interacting susceptibility and the interaction potential. The interaction is typically modelled by a repulsive Coulomb interaction $V(\mathbf{q})|_{c}$ for the charge susceptibility. The full charge susceptibility $\chi^{RPA}_{\mathbf{c}}$ is given by the RPA summation as:
\begin{equation}
    \chi^{RPA}_{ij}(\tau)|_{\mathbf{c}} = \chi_{0\,\,ij}(\tau) + V(\mathbf{q})|_{c} \sum_{k} \chi_{0\,\,ik}(\tau) \chi^{RPA}_{kj}(\tau)|_{\mathbf{c}}
\end{equation}
Here, $\chi_{0\,\,ij}(\tau)$ is the non-interacting Lindhard susceptibility calculated earlier, and $V(\mathbf{q})|_{c}$ is the on-site Coulomb interaction.

In momentum space, this can be written as:
\begin{equation}
    \chi^{RPA}(\mathbf{q}, i\omega_n)|_{\mathbf{c}} = \frac{\chi_{0}(\mathbf{q}, i\omega_n)}{1 - V(\mathbf{q})|_{c} \chi_{0}(\mathbf{q}, i\omega_n)}
\end{equation}
where $\mathbf{q}$ is the momentum and $i\omega_n$ is the Matsubara frequency. This expression describes how the interaction $V(\mathbf{q})|_{c}$ enhances the charge fluctuations. For full derivation, see Chapter 9. of \cite{BrussandFlensberg}. It is to be noted that in the spin channel, the $V(\mathbf{q})|_{s}$ corresponds to a Heisenberg-type of exchange interaction instead of Coulomb interaction with a different sign.

\section{Parameters}\label{App:Parameters}
\begin{table}[!htbp]
\centering
\resizebox{\textwidth}{!}{
\begin{tabular}{|cccccc|}
\hline
\multicolumn{6}{|c|}{\textbf{Fits on \textit{ab-initio} results:}} \\ \hline
\multicolumn{1}{|c|}{\textbf{Compound}} & \multicolumn{1}{c|}{\textbf{Ref}} & \multicolumn{1}{c|}{\textbf{Orbitals}} & \multicolumn{1}{c|}{\textbf{On-Site energy (\,eV)}} & \multicolumn{1}{c|}{\textbf{NN hopping (\,eV)}} & \textbf{NNN hopping (\,eV)} \\ \hline
\multicolumn{1}{|c|}{$LaNiO_2$} & \multicolumn{1}{c|}{\cite{PhysRevX.10.011024}} & \multicolumn{1}{c|}{$Ni-d_{x^2-y^2}$} & \multicolumn{1}{c|}{$0.249$} & \multicolumn{1}{c|}{$-0.368$} & $0.099$ \\ \hline
\multicolumn{1}{|c|}{$LaNiO_2$} & \multicolumn{1}{c|}{\cite{Lee2004Oct}} & \multicolumn{1}{c|}{$Ni-d_{x^2-y^2}$} & \multicolumn{1}{c|}{$0.093$} & \multicolumn{1}{c|}{$-0.381$} & $0.081$ \\ \hline
\multicolumn{1}{|c|}{\multirow{2}{*}{$LaNiO_2$}} & \multicolumn{1}{c|}{\multirow{2}{*}{\cite{Plienbumrung2022Oct}}} & \multicolumn{1}{c|}{$Ni-d_{x^2-y^2}$} & \multicolumn{1}{c|}{$0.281$} & \multicolumn{1}{c|}{$-0.380$} & $0.088$ \\ \cline{3-6} 
\multicolumn{1}{|c|}{} & \multicolumn{1}{c|}{} & \multicolumn{1}{c|}{$Interstitial: s$} & \multicolumn{1}{c|}{$1.493$} & \multicolumn{1}{c|}{$-0.031$} & $-0.111$ \\ \hline
\multicolumn{1}{|c|}{$LaNiO_2$} & \multicolumn{1}{c|}{\cite{Klett2022Feb}} & \multicolumn{1}{c|}{$Ni-d_{x^2-y^2}$} & \multicolumn{1}{c|}{--} & \multicolumn{1}{c|}{$0.395$} & $-0.095$ \\ \hline
\multicolumn{1}{|c|}{$NdNiO_2$} & \multicolumn{1}{c|}{\cite{Zhang2020Feb}} & \multicolumn{1}{c|}{$Ni-d_{x^2-y^2} + O-p_{x/y}$} & \multicolumn{1}{c|}{--} & \multicolumn{1}{c|}{$0.254$} & $-0.030$ \\ \hline
\multicolumn{1}{|c|}{$LaNiO_2$} & \multicolumn{1}{c|}{\multirow{2}{*}{\cite{Kitatani2020Aug}}} & \multicolumn{1}{c|}{$Ni-d_{x^2-y^2}$} & \multicolumn{1}{c|}{$0.2689$} & \multicolumn{1}{c|}{$-0.3894$} & $0.0977$ \\ \cline{1-1} \cline{3-6} 
\multicolumn{1}{|c|}{$NdNiO_2$} & \multicolumn{1}{c|}{} & \multicolumn{1}{c|}{$Ni-d_{x^2-y^2}$} & \multicolumn{1}{c|}{$0.2502$} & \multicolumn{1}{c|}{$-0.3974$} & $0.0933$ \\ \hline
\multicolumn{1}{|c|}{\multirow{2}{*}{$LaNiO_2$}} & \multicolumn{1}{c|}{\multirow{2}{*}{\cite{Hepting2020Apr}}} & \multicolumn{1}{c|}{$Ni-d_{x^2-y^2}$} & \multicolumn{1}{c|}{$0.267$} & \multicolumn{1}{c|}{$-0.355$} & -- \\ \cline{3-6} 
\multicolumn{1}{|c|}{} & \multicolumn{1}{c|}{} & \multicolumn{1}{c|}{$La-d_{z^2}$} & \multicolumn{1}{c|}{$1.132$} & \multicolumn{1}{c|}{$-0.164$} & -- \\ \hline
\end{tabular}
}
\caption{}
\label{ParameterTable1}
\end{table}

\begin{table}[!htbp]
\centering
\resizebox{\textwidth}{!}{
\begin{tabular}{|c|c|c|c|c|c|}
\hline
\textbf{Compound} & \textbf{Ref} & \textbf{Orbital} & \textbf{NN-hopping (\,eV)} & \textbf{NNN-hopping (\,eV)} & \textbf{$\Delta_{d_{x^2-y^2} -d_{xy}}$ (\,eV)} \\ \hline
\multirow{2}{*}{$LaNiO_2$} & \multirow{6}{*}{\cite{Xie2022Jul}} & $Ni-d_{x^2-y^2} $ & $-0.370$ & $0.10$ & \multirow{2}{*}{$1.39$} \\ \cline{3-5}
 &  & $Ni-d_{xy}$ & $-0.16$ & $-0.05$ &  \\ \cline{1-1} \cline{3-6} 
\multirow{2}{*}{$PrNiO_2$} &  & $Ni-d_{x^2-y^2} $ & $-0.370$ & $0.10$ & \multirow{2}{*}{$1.41$} \\ \cline{3-5}
 &  & $Ni-d_{xy}$ & $-0.16$ & $-0.05$ &  \\ \cline{1-1} \cline{3-6} 
\multirow{2}{*}{$NdNiO_2$} &  & $Ni-d_{x^2-y^2} $ & $-0.370$ & $0.10$ & \multirow{2}{*}{$1.42$} \\ \cline{3-5}
 &  & $Ni-d_{xy}$ & $-0.16$ & $-0.05$ &  \\ \hline
\end{tabular}
}
\caption{}
\label{ParameterTable2}
\end{table}

\begin{table}[!htbp]
\centering
\begin{tabular}{|c|c|ccccc|}
\hline
\textbf{Compound} & \textbf{Ref.} & \multicolumn{5}{c|}{\textbf{Parameters (\,eV)}} \\ \hline
\multirow{8}{*}{$LaNiO_2$} & \multirow{2}{*}{\cite{PhysRevX.10.011024}} & \multicolumn{1}{c|}{$\Tilde{\varepsilon}_{2p}$} & \multicolumn{1}{c|}{$\Tilde{\varepsilon}_{3d_{x^2-y^2}}$} & \multicolumn{1}{c|}{$\Tilde{\varepsilon}_{3d_{z^2}}$} & \multicolumn{1}{c|}{$t_{pd}$} & $t^{\prime}_{pd}$ \\ \cline{3-7}
 &  & \multicolumn{1}{c|}{$-5.41$} & \multicolumn{1}{c|}{$-1.02$} & \multicolumn{1}{c|}{$-1.73$} & \multicolumn{1}{c|}{$-1.23$} & $0.20$ \\ \cline{2-7}
 
 & \multirow{4}{*}{\cite{Hepting2020Apr}} & \multicolumn{1}{c|}{$\Tilde{\varepsilon}_{2p}$} & \multicolumn{1}{c|}{$\Tilde{\varepsilon}_{3d_{x^2-y^2}}$} & \multicolumn{1}{c|}{$\Tilde{\varepsilon}_{3d_{z^2}}$} & \multicolumn{1}{c|}{$t_{pd}$} & $t^{\prime}_{pd}$ \\ \cline{3-7} 
  &  & \multicolumn{1}{c|}{$-3.26$} & \multicolumn{1}{c|}{$0.70$} & \multicolumn{1}{c|}{$0.04$} & \multicolumn{1}{c|}{$-1.20$} & $-0.30$ \\ \cline{3-7}
  
 & & \multicolumn{1}{c|}{$\Tilde{\varepsilon}_{5d_z^2}$} & \multicolumn{1}{c|}{$t^{\prime}_{dd}$} & \multicolumn{1}{c|}{$t^{\prime\prime}_{dd}$} & \multicolumn{1}{c|}{$t_{dd}$} & \multicolumn{1}{c|}{$--$}   \\ \cline{3-7} 

 &  & \multicolumn{1}{c|}{$-2.42$} & \multicolumn{1}{c|}{$-0.5$} & \multicolumn{1}{c|}{$--$} & \multicolumn{1}{c|}{$--$} & $--$ \\ \cline{2-7}
  
 & \multirow{2}{*}{\cite{Jiang2019Nov}} & \multicolumn{1}{c|}{$\Delta_{dp}$} & \multicolumn{1}{c|}{$\Tilde{\varepsilon}_{3d_{x^2-y^2}}$} & \multicolumn{1}{c|}{$\Tilde{\varepsilon}_{3d_{z^2}}$} & \multicolumn{1}{c|}{$\Tilde{\varepsilon}_{5d_{z^2}}$} & $t^{\prime}_{dd}$ \\ \cline{3-7} 
 &  & \multicolumn{1}{c|}{$-6$} & \multicolumn{1}{c|}{$0.40$} & \multicolumn{1}{c|}{$-2.94$} & \multicolumn{1}{c|}{$2.649$} & $0.253$ \\ 
 \hline
\multirow{2}{*}{$NdNiO_2$} & \multirow{2}{*}{\cite{Zhang2020Feb}} & \multicolumn{1}{c|}{$\Delta_{dp}$} & \multicolumn{1}{c|}{$\Tilde{\varepsilon}_{3d_{x^2-y^2}}$} & \multicolumn{1}{c|}{$t_{pd}$} & \multicolumn{1}{c|}{$t_{pp}$} & -- \\ \cline{3-7} 
 &  & \multicolumn{1}{c|}{--} & \multicolumn{1}{c|}{$4.2$} & \multicolumn{1}{c|}{$1.3$} & \multicolumn{1}{c|}{$0.6$} & -- \\ \hline
\end{tabular}

\caption{}
\label{ParameterTable3}
\end{table}

\begin{table}[!htbp]
\centering
\begin{tabular}{|c|c|c|c|}
\hline
\textbf{Compound} & \textbf{Ref} & $\Delta_{dp} $ (\,eV) & $t_{pd}$ (\,eV)\\ \hline
$LaNiO_3$ & \cite{Lopez-Bezanilla2019Jan} & $4.4 $ & $-1.3$ \\ \hline
\end{tabular}
\caption{}
\label{ParameterTable4}
\end{table}

\section*{References}
\bibliographystyle{unsrt}
\bibliography{References}

\end{document}